\documentclass[10pt]{asme2ej}

\usepackage{epsfig} %% for loading postscript figures

\usepackage{color}
\usepackage{amssymb}
\usepackage{amsmath}
\usepackage{float}
\usepackage{url}
\newcommand{\bx}{\mathbf{x}}

\newcommand{\bff}{\mathbf{f}}

\newcommand{\bw}{\mathbf{w}}

\newcommand{\g}{\mathbf{\mathtt{g}}}

\newcommand{\bI}{\mathbf{I}}
\newcommand{\bA}{\mathbf{A}}
\newcommand{\bV}{\mathbf{V}}
\newcommand{\bU}{\mathbf{U}}
\newcommand{\bC}{\mathbf{C}}

\newcommand{\bW}{\mathbf{W}}
\newcommand{\bK}{\mathbf{K}}

\newcommand{\bxi}{\boldsymbol{\xi}}

\newcommand{\btheta}{\boldsymbol{\theta}}

\newcommand{\balpha}{\boldsymbol{\alpha}}
\newcommand{\bbeta}{\boldsymbol{\beta}}

\newcommand{\bfeta}{\boldsymbol{\eta}}

\newcommand{\bLam}{\boldsymbol{\Lambda}}

\newcommand{\calF}{\mathcal{F}}

\newcommand{\calJ}{\mathcal{J}}

\newcommand{\calU}{\mathcal{U}}

\newcommand{\R}{\mathbb{R}}
\newcommand{\N}{\mathbb{N}}
\newcommand{\E}{\mathbb{E}}
\newcommand{\Prob}{\mathbb{P}}

\usepackage[ruled]{algorithm2e}

\title{Gradient-informed basis adaptation for Legendre Chaos expansions}

\author{Panagiotis A. Tsilifis
	\affiliation{
	CSQI, Institute of Mathematics\\
	School of Basic Sciences\\
    \'{E}cole Polytechnique F\'{e}d\'{e}rale de Lausanne\\
    CH-1015 Lausanne, Switzerland\\
    Email: panagiotis.tsilifis@epfl.ch
    }
}

\begin{document}

\maketitle    

%%%%%%%%%%%%%%%%%%%%%%%%%%%%%%%%%%%%%%%%%%%%%%%%%%%%%%%%%%%%%%%%%%%%%%
\begin{abstract}
The recently introduced basis adaptation method for Homogeneous (Wiener) Chaos expansions is explored in a new context where the rotation/projection matrices are computed by discovering the active subspace where the random input exhibits most of its variability. In the case where a 1-dimensional active subspace exists, the methodology can be applicable to generalized Polynomial Chaos expansions, thus enabling the projection of a high dimensional input to a single input variable and the efficient estimation of a univariate chaos expansion. Attractive features of this approach, such as the significant computational savings and the high accuracy in computing statistics of interest are investigated.
\end{abstract}

%%%%%%%%%%%%%%%%%%%%%%%%%%%%%%%%%%%%%%%%%%%%%%%%%%%%%%%%%%%%%%%%%%%%%%
%\begin{nomenclature}
%\entry{A}{You may include nomenclature here.}
%\entry{$\alpha$}{There are two arguments for each entry of the nomemclature environment, the symbol and the definition.}
%\end{nomenclature}

%The primary text heading is  boldface and flushed left with the left margin.  The spacing between the  text and the heading is two line spaces.

%%%%%%%%%%%%%%%%%%%%%%%%%%%%%%%%%%%%%%%%%%%%%%%%%%%%%%%%%%%%%%%%%%%%%%

\section{Introduction}

% Computarional challenges in UQ problems. High dimensionality
While new technological advancements and the rapid increase of computational power simply seem to motivate the need for solving even more complex physical problems, uncertainty quantification (UQ) tasks seem to suffer from the never-ending challenge of the ``relatively" limited computational resources available to the experimenter. To overcome this challenge of developing problem- and computer code-independent efficient methodologies that will enable experimentation and predictive capabilities, one necessarily focuses on further advancing the mathematical and algorithmic tools for quantifying the various forms of uncertainties and their effects on physical models.

% Computer surrogates
Most standard UQ approaches involve the use of Monte Carlo (MC) methods \cite{robert_casella} where several samples are drawn according to the model input distribution and their corresponding outputs are used to compute certain statistics of interest. The use of such methods typically varies from uncertainty propagation problems \cite{morokoff} to calibration problems \cite{tarantola, mosegaard} to stochastic optimization \cite{spall_book, spall_alg} and experimental design problems \cite{huan_marzouk,tsilifis_design}. The slow convergence of the method along with the usually expensive computational models can easily make the approach unaffordable and infeasible, causing one to resort to cheaper alternatives. These consist of substituting the computational model with a surrogate that can be repeatedly evaluated at almost no cost, therefore accelerating computations. Such surrogates can be, for instance, polynomial chaos expansions (PCE) \cite{ghanem_sfem, xiu_askey, babuska, reagan_nonintrusive}, Gaussian Processes \cite{bilionis_multi, bilionis_separ} or adaptive sparse grid collocation \cite{ma}. The common characteristic in all the above constructions is that an initial set of forward evaluations at preselected points is used to construct a functional approximation of the original computational model. After such an approximation becomes available, one no longer relies on the expensive-to-evalute computer code but instead uses the surrogate. Although this characteristic generally outperforms MC methods, it can still fail to address the issue of computational efficiency due to the \emph{curse of dimensionality} effect, that is the increase of the dimensionality of the uncertain input results is an exponentially increasing number of evaluations required to compute the surrogate parameters. 

% Dimensionality reduction
Several ways to achieve dimensionality reduction have been proposed in the literature, among which are sensitivity analysis methods \cite{saltelli} that rank the inputs according to their influence, thus allowing to neglect the components with respect to which, the output is insensitive. These methods involve for example variance decomposition methods such as the Sobol' indices \cite{sobol,owen}. Similar approaches might consist of spectral methods, such as the Karhunen-Lo\'{e}ve expansion (KLE) \cite{karhunen,loeve,ghanem_sfem} for random vectors and random fields, where one expands in a series of scalar variables, the importance of which is determined by the eigenvalues of the covariance matrix. The series can be truncated accordingly, leaving out the unimportant scales of fluctuation of the random quantity. For an empirical version of the KLE one might also consider the Principal Component Analysis (PCA) \cite{pearson_pca} or the kernel PCA \cite{ma_pca}. At last, one can proceed with the active subspace (AS) method \cite{constantine, constantine_scram, lukaczyk} of discovering the low-dimensional projection of the random input space where the model exhibits maximal variability. This makes use of a spectral decomposition of the covariance of the gradient vector, the eigenvectors of which reveal the projection mapping to the active subspace. % Active subspaces 

% Polynomial Chaos 
In this work, our attention is focused on reduction methods that are applicable on Polynomial Chaos expansions. PCE's provide an analytic and compact representation of the model output in terms of a number of scalar random variables, typically referred to as the germs, through polynomials that are pairwise orthogonal with respect to inner product (expectation) of the underlying Hilbert space of square integrable random variables. Their use, pioneered by \cite{ghanem_sfem}, has been explored throughout several contexts and applications such as flow through heterogeneous porous media \cite{ghanem_wrr, ghanem_dham}, fluid dynamics \cite{najm, le_maitre_etal, xiu_nonintrusive} aeroelasticity \cite{arnst_ghanem} etc. for either forward propagation \cite{ghanem_doostan_redhorse, ghanem_redhorse} or inference problems \cite{marzouk_etal, marzouk_najm, ghanem_doostan}.

% Sobol' indices and basis adaptation in PCEs
Most of the aforementioned dimensionality reduction methods can be applied on PCE's once the expansion becomes known. In \cite{sudret,crestaux}, explicit formulas for the Sobol' sensitivity indices were derived with respect to the coefficients of chaos expansions, thus enabling the fast computation of variance decomposition factors whenever such expansions are available. The idea was used in the context of stochastic differential equations \cite{le_maitre_knio} to separate the influence of uncertainty due to random paramers from that caused by the driving white noise and in a similar fashion it has also been applied to large-eddy simulations \cite{lucor} for sensitivity analysis due to turbulent effects. However, the key challenge of reducing the dimensionality of the input, in order to make the computation of the coefficients a feasible task still remains unsolved and most of the works still focus their attention on simply obtaining sparse representations by adaptive finite elements \cite{blatman}, $\ell$1-minimization \cite{hampton,yang_karn} and compressive sensing \cite{hampton_doostan}.

Recently \cite{tipireddy} it was observed that for Hermite expansions with Gaussian input variables, the original input can be rotated using isometries onto the underlying Gaussian Hilbert space to obtain expansions with respect to the new basis that concentrate their dependence on only a few components. Certain choices for the rotation matrix were analytically seen not only to guarantee certain levels of sparsity in the expansion but even concentrate the probabilistic behavior of scalar quantities of interest (QoI) on low-dimensional Gaussian subspaces. Computational schemes for efficiently computing the coefficients of the adapted expansion were also developed and the numerical results were impressive. The basis adaptation concept was further extended from scalar QoI's to random fields \cite{tsilifis_adapt}, where explicit formulas for computing the coefficients with respect to any rotation were derived and the case of a parametric family of rotations was discussed that gives rise to expansions with a Gaussian process input. Even though the idea of rotating the basis has been further applied to design optimization problems \cite{thimmisetty} and has been used to develop efficient schemes coupled with compressing sensing methods \cite{yang_lin, tsilifis_CS}, it is still quite restricted to the Homogeneous Chaos expansions, where the Gaussian distribution remains invariant under rotations, a property that is not valid in other cases.

The key difference when applying the above idea in random vectors other than Gaussians, is that the distribution or the rotated variables is, in general, not analytically available and the construction of polynomials, orthogonal with respect to this distribution is not an easy task. In this paper we are exploring the special case where an isometry can be used to rotate the basis in the same fashion as in the basis adaptation methodology, such that only a 1-dimensional component of the new variables will suffice to build an accurate polynomial chaos expansion. In this, quite restrictive case, the input can easily be mapped to a uniform distribution via its own cummulative distribution function and therefore the Legendre polynomials can be used to expand the series. To find the proper rotation matrix, we make use of the active subspace methodology that successfully discovers the rotation such that maximal variability can be achieved in one only variable. To do so, unlike the traditional active subspace method where the covariance matrix of the gradient vector is computed via MC methods, we make use of an initial chaos expansion that is obtained using low-order quadratures, that is with very few model evaluations. In this case the matrix can be analytically computed at no additional cost. Once the active subspace is revealed, a new one-dimensional Legendre chaos expansion of relatively high order can be easily constructed as again, it does not demand many evaluations. Apart from the fact that the approach enables surrogate construction of expensive models that would be otherwise infeasible, we also discover some attractive features: The capability of obtaining high order one-dimensional expansions allows for very accurate predictions of the probabilistic behavior of QoI's where a lower order full-dimensional expansion fails. 

This paper is structured as follows: Section \ref{sec:metho} reviews the polynomial chaos and active subspace approaches and then presents how the first can be used to analytically compute the latter along with an algorithm for efficient construction of 1d Legendre Chaos expansions. In Section \ref{sec:examples} we apply the methodology to two numerical examples. First, a simple quadratic polynomial function with analytically known active subspace is used for validation and second, a multiphase flow problem that simulates transport of ammonium and its oxidation to nitrite along a 1-dimensional rod is used to draw my conclusions.

%We apply our methodology in two applications. First, the multiphase transport and oxidation of ammonium along a one-dimensional rod, where certain decay effects and material properties are assumed random and the concentration of ammonium can be hard to predict near points where it is almost zero. Second, the multiphase turbulent large-eddy simulation of a HIFire SCRAMJET combustor.

\section{Methodology}
\label{sec:metho}

To provide a proper setting upon which our methodology is developed, we
first consider a general response function $f : D \subset \R^d \to \R$
where the points $\bxi \in D$ will be thought of as \emph{inputs}
and their mappings $f(\bxi)$ will the the \emph{model ouputs} or the
\emph{quantity of interest (QoI)}. Typically $D$ is also equipped with some probability measure
$\Prob$ therefore its elements are $D$-valued random variables from a
space of events $\Omega$ to $D$. A $\sigma$-algebra $\calF$ that
consists of the $\Prob$-measurable subsets of $\Omega$ or in other words,
the inverse mappings of the Borel subsets of $D$ is naturally defined
and thus the probability triplet $(\Omega, \calF, \Prob)$ is the space
on which we will be working. 

\subsection{Generalized Polynomial Chaos}

Throughout this paper we will assume that the quantity $f(\bxi)$ has
finite variance and therefore is a square integrable function, that is
$f \in L^2(\Omega, \calF, \Prob)$. It is known then \cite{xiu_askey}, that $f$ admits a
series expansion in terms of orthogonal polynomials of $\bxi$, given as 
\begin{equation}
\label{eq:chaos_expansion}
f(\bxi) = \sum_{\balpha, |\alpha| = 0}^{\infty} f_{\balpha} \psi_{\balpha}(\bxi),
\end{equation}
where $\balpha = (\alpha_1, \dots, \alpha_d) \in \calJ := \N \cup
\{\mathbf{0}\}$ are finite-dimensional multiindices with norm
$|\balpha| = \alpha_1 + \dots + \alpha_d$ and $\psi_{\balpha}$ are
$d$-dimensional polynomial functions of $\bxi$ that are orthogonal
with respect to the measure defined by the density function $p(\bxi)$,
that is 
\begin{equation}
\E\left\{\psi_{\balpha}(\bxi)\psi_{\bbeta}(\bxi)\right\} = \int_{D}
\psi_{\balpha}(\bxi)\psi_{\bbeta}(\bxi) p(\bxi) d\bxi =
||\psi_{\alpha}||^2 \delta_{\balpha, \bbeta}
\end{equation}
where $||\psi_{\balpha}|| =
\left(\E\{\psi_{\balpha}(\bxi)^2\}\right)^{1/2}$ and
$\delta_{\balpha,\bbeta}$ is the Dirac delta function which is $1$ if
$\balpha = \bbeta$ and $0$ otherwise. Without loss of generality we
assume here that $||\psi_{\balpha}|| = 1$, that is the polynomials are
normalized. We will refer to eq. (\ref{eq:chaos_expansion}) as the \emph{polynomial
  chaos (PC) expansion} of $f$. In the case where $\bxi = (\xi_1,
\dots, \xi_d)$ consists of independent and identically distributed
random variables, the polynomials are expanded as 
\begin{equation}
\psi_{\balpha} = \prod_{i = 1}^d \psi_{\alpha_i}(\xi_i)
\end{equation}
where $\psi_{\alpha_i}(\xi_i)$ are univariate polynomials of $\xi_i$ of
order $\alpha_i$, $i = 1, \dots, d$. Common choices of the
density function $p(\bxi) = \prod_{i = 1}^d p(\xi_i)$ give rise to known forms of the
polynomials, for instance the Gaussian, Uniform, Gamma and Beta
distributions correspond to Hermite, Legendre, Laguerre and Jacobi polynomials
respectively \cite{xiu_askey}.

In practice, we work with truncated versions of
(\ref{eq:chaos_expansion}), that is for $Q \in \N$, $\calJ_Q :=
\left\{\balpha \in \calJ:\ |\balpha|\leq Q\right\}$, we will assume
that $f$ can be \emph{accurately} approximated by
\begin{equation}
\label{eq:chaos_trunc}
f(\bxi) \approx \sum_{\balpha\in \calJ_Q} f_{\balpha}\psi_{\balpha}(\bxi).
\end{equation}
The above expansion consists of 
\begin{equation}
N_Q = \left(\begin{array}{c}d + Q \\ Q\end{array}\right) = \frac{(d + Q)!}{d! Q!}
\end{equation}
basis terms and the estimation of the corresponding coefficients is
typically a challenging task. Several approaches have been developed
in the literature for the estimation of the coefficients
$\{f_{\balpha}\}_{\balpha\in \calJ_Q}$, divided in two main
categories: Intrusive and non-intrusive methods. The first, pioneered by \cite{ghanem_sfem}, treats the solution of a differential equation as a random field that can be written as in (\ref{eq:chaos_trunc}), where the coefficients vary as functions of the spatial or time parameters associated with the computational domain. The expression is then
substituted in the equation satisfied by $f$, in order to derive the governing equations satisfied by the coefficients, which then need to be solved once. Non-intrusive
approaches involve estimation of the coefficients using the relation
\begin{equation}
\label{eq:projection_coeff}
f_{\balpha} = \E\left\{f(\bxi)\psi_{\balpha}(\bxi)\right\}
\end{equation}
which is computed using numerical integration techniques. As mentioned in the previous section, both
approaches suffer by the curse of dimensionality which in the first
case implies that the system of equations to be solved increases
exponentially as a function of the dimensionality $d$ and in the
second case, the exponential increase is on the number of collocation
points where the model $f$ needs to be evaluated. 

\subsection{Active subspaces}

\subsubsection{Discovering the active subspace}

Let us assume that the function $f$ can be well-approximated by 
\begin{equation}
\label{eq:link_func}
f(\bxi) \approx g(\bW^T\bxi)
\end{equation}
where $\bW$ is a $d\times d'$ matrix with orthonormal columns, that is
it satisfies 
\begin{equation}
\bW^T\bW = \bI_{d'}
\end{equation}
and $g: \R^{d'} \to \R$ is a $d'$-dimensional function that will be
called the \emph{link} finction. Intuitively, such an assumption
implies that the domain $D$ can be rotated using a $d\times d$ orthonormal matrix
\begin{equation}
\label{eq:A_decomp}
\bA = \left[\begin{array}{cc}\bW & \bV\end{array} \right]
\end{equation} 
with $\bA^T\bA = \bI_d$ and $\bW$, $\bV$ being $d\times d'$ and $d\times (d-d')$ matrices
respectively such that $f$ exhibits most of its variation on the space
spanned by $\{\bW^T\bxi :\ \bxi \in D \}$ while it remains insensitive to variations
on its orthogonal complement, thus motivating the term \emph{active
  subspace} \cite{constantine}.

As it is clear from the above, the main challenge in the above
construction is the determination of the
\emph{rotation} matrix $\bA$ such that the active subspace defined by
$\bW$ will have the minimum possible dimensionality $d'$. To explore
the directions along which $f$ exhibits greatest sensitivity, we
define the $d \times d$ matrix 
\begin{equation}
\bC = \E\left\{\nabla f(\bxi) \nabla f(\bxi)^T \right\},
\end{equation}
where $\nabla f(\bxi) = \left(\frac{\partial f(\bxi)}{\partial \xi_1},
  \dots, \frac{\partial f(\bxi)}{\partial \xi_d}\right)^T$ is the
gradient vector of $f$. This is essentially the uncentered covariance
matrix of $\nabla f(\bxi)$ and is a symmetric and positive
semi-definite matrix, therefore it admits a decomposition as 
\begin{equation}
\bC = \bU \bLam \bU^T,
\end{equation}
where $\bLam$ is the diagonal matrix with entries the eigenvalues of
$\bC$ in decreasing order, namely 
$\lambda_i$, $i = 1, \dots, d$ with $\lambda_1 \geq \dots \geq \lambda_d$ and $\bU$ is the
unitary matrix whose $i$th column is the eigenvector of $\bC$
corresponding to $\lambda_i$. By setting $\bA = \bU$ and $\bfeta = \bA^T\bxi$ it can be shown
\cite{constantine} that 
\begin{equation}
\E\left\{\nabla_{\bfeta} f)^T(\nabla_{\bfeta} f)\right\} = \lambda_1 + \cdots + \lambda_d
\end{equation}
and by decomposing $\bA$ as in (\ref{eq:A_decomp}) and setting $\bfeta_{\bw} =
\bW^T\bxi$ gives us that the mean-squared gradients of $f$ with
respect to $\bfeta_{\bw}$ will be the sum of the $d'$ largest
eigenvalues. This construction provides us a way of rotating the input
space $D$ and separating it into two subspaces namely the active
subspace and its orthogonal complement by simply selecting the $d'$ most
dominant eigenvalues of $\bC$ and their corresponding
eigenvectors. Then the function $f$ can be approximated as in eq. \ref{eq:link_func}.

\subsubsection{Active subspace computation for PC expansions and reduction to univariate Legendre Chaos}

The key challenge in the active subspace methodology is
the computation of the \emph{gradient matrix} $\bC$, the standard way
of which is by estimating the expectation using Monte Carlo sampling. Recently a
gradient-free approach for determining the projection matrix when $f$
is approximated by a Gaussian Process was developed
\cite{tripathy}. For this purpose, assume that a PC expansion is available for $f$ as in
eq. (\ref{eq:chaos_trunc}). Then $\bC$ can be written as 
\begin{equation}
\bC \approx \E\left\{ \nabla \left(\sum
f_{\balpha} \psi_{\balpha}(\bxi)\right) \cdot \nabla \left(\sum
f_{\bbeta} \psi_{\bbeta}(\bxi) \right) ^T\right\}
\end{equation}
and the entries $C_{ij}$ will be given by 
\begin{eqnarray}
\label{eq:gradient_entries}
\begin{array}{ccl}
C_{ij} & \approx & \displaystyle{ \E\left\{ \frac{\partial}{ \partial \xi_i} \left(\sum
f_{\balpha} \psi_{\balpha}(\bxi)\right) \cdot \frac{\partial}{\partial
\xi_j} \left(\sum
f_{\bbeta} \psi_{\bbeta}(\bxi) \right) ^T\right\}} \\ & = &
\displaystyle{ \E\left\{ \left(\sum
f_{\balpha} \frac{\partial \psi_{\balpha}(\bxi)}{ \partial \xi_i} \right) \cdot \left(\sum
f_{\bbeta} \frac{\partial \psi_{\bbeta}(\bxi)}{\partial \xi_j} \right)
^T\right\} } \\ & = & 
\displaystyle{ \sum_{\balpha} \sum_{\bbeta} f_{\balpha} f_{\bbeta}
  \E\left\{ \frac{\partial \psi_{\balpha}(\bxi)}{ \partial \xi_i} \cdot
  \frac{\partial \psi_{\bbeta}(\bxi)}{\partial \xi_j}\right\} }\\ & = &
\displaystyle{\bff^T \bK_{ij} \bff
}
\end{array}
\end{eqnarray}
where $\bK_{ij}$ is the \emph{stiffness} matrix with entries
\begin{equation}
\label{eq:stiffness}
\left(\bK_{ij} \right)_{\balpha\bbeta} = \E\left\{ \frac{\partial \psi_{\balpha}(\bxi)}{ \partial \xi_i} \cdot
  \frac{\partial \psi_{\bbeta}(\bxi)}{\partial \xi_j}\right\}
\end{equation}
and $\bff = \{f_{\balpha}\}_{\balpha\in \calJ_Q}$ is the vectorized
representation of the chaos coefficients. The values of the entries of the
stiffness matrices $\bK_{ij}$, $i,j = 1,\dots, d$ depend solely on the
polynomials used in the PC expansion and their corresponding
probability measures with respect to which the expectation is taken
and can be computed independently of the nature of
the function $f$. Detailed computation of $\bK_{ij}$ of the case of Legendre polynomials is provided in Appendix \ref{sec:stiff_comp}. Computation of $\bC$ generally requires 
full knowledge of the coefficients $\bff$ and can be an expensive
task. However, as we describe in the next section, a relatively \emph{cheap} estimation
of the coefficients, based on low level quadrature rules can suffice
of this purpose.

Let us now assume that the matrix $\bC$ is available and that $\lambda_1 \gg
\lambda_2$. This implies that the rotation matrix $\bA$ can be
decomposed in a 1-dimensional vector $\bw$ with $\bw^T\bw = 1$ and a
$d\times (d-1)$ matrix $\bV$ such that $f$ exhibits most of its
variation on the space spanned by $\eta = \bw^T\bxi$ and can be
accurately approximated as a function of $\eta$ as in eq. (\ref{eq:link_func}).
The square integrability condition of $f$ directly implies that $f$ can
be expanded in a polynomial chaos expansion in terms of
polynomials that are orthogonal with respect to the probability measure
of $\eta$. To avoid the structure of such polynomials, since the
probability measure $p(\eta)d\eta$ can be arbitrarily complex, we
introduce the uniform $\calU(-1,1)$ germ 
\begin{equation}
\zeta = 2F_{\eta}(\eta) - 1,
\end{equation}
where $F_{\eta}(\cdot)$ is the cummulative probability distribution of
$\eta$ and we write 
\begin{equation}
f(\bxi) \approx g(\eta) = g\left(F_{\eta}^{-1}\left(\frac{\zeta+
      1}{2}\right)\right) := \g(\zeta).
\end{equation}
Finally $g$ can be expanded as 
\begin{equation}
\label{eq:1d_pce}
\g(\zeta) = \sum_{n= 0}^{N_\zeta} \g_{n} \psi_n(\zeta)
\end{equation}
where $\psi_n(\zeta)$ are the normalized univariate Legendre
polynomials and thus we can achieve a 1-dimensional chaos
decomposition of $f$.

\subsection{Efficient basis reduction using pseudo-spectral projections}

Using eq. (\ref{eq:projection_coeff}) the coefficients of the chaos
expansion of $f$ can be estimated after approximating the integral
with 
\begin{equation}
\label{eq:full_quadr}
f_{\balpha} = \int_{D} f(\bxi) \psi_{\balpha}(\bxi)p(\bxi)d\bxi
\approx \sum_{i=1}^s f(\bxi^{(i)})\psi_{\balpha}(\bxi^{(i)}) w_i, \ \
\balpha \in \calJ_Q,
\end{equation}
using a quadrature rule $\{\bxi^{(i)}, w_i\}_{i=1}^{s_d}$, where
$\{\bxi^{(i)}\}_{i=1}^{s_d}$ are quadrature points and
$\{w_i\}_{i=1}^{s_d}$ are the corresponding weights.
As mentioned above, such a procedure can be prohibitive for relatively
large $d$ or for computationally expensive models $f$ that exhibit
high nonlinearity. It is desirable, in such cases, to develop a
computational strategy that reduces the computational resources by
limiting the number of model evaluations. Provided that a
1-dimensional active subspace exists, writing $f$ in the form
(\ref{eq:1d_pce}) and computing the coefficients
$\{\g_n\}_{n=0}^{N_\zeta}$ would be an efficient alternative. Since
this requires the knowledge of the projection vector $\bw$, one could
perform the following steps: First, a low level quadrature rule that
consists of an affordable number of points can be performed in order
to compute the low order coefficients $\{f_{\balpha}\}_{\balpha \in
  \calJ_{Q_0}}$, $Q_0 < Q$ and an estimate of $\bC$ and therefore of
$\bw$ can be obtained. Then the PC coefficients of $\g$ can be
computed in a similar manner by 
\begin{equation}
\g_n \approx \sum_{i = 1}^{s_1} \g(\zeta^{(i)}) \psi_{n}(\zeta^{(i)}) w_i
\end{equation}
where $\{\zeta^{(i)}\}_{i=1}^{s_1}$ are $1$-dimensional quadrature points on the $[-1,1]$ space. In
practice, evaluation of $\g(\zeta^{(i)})$ will require transforming
$\zeta^{(i)}$ to $\eta^{(i)} = F_{\eta}^{-1}(\frac{\zeta^{(i)} + 1}{2})$ and
$\bfeta^{(i)} = \bA\bxi^{(i)}$ where $\bfeta^{(i)}$ is a $d$-dimensional arbitrary
completion of $\eta^{(i)}$, where the remaining $d-1$ entries will essentially
play no role in the model output as they span the subspace with
respect to which $f$ is approximately invariant. This procedure is
summarized in the pseudo-algorithm described in Algorithm \ref{alg:quadr}.

\begin{algorithm}[h]
\caption{Non intrusive implementation \label{alg:quadr}}
\SetKwInOut{Input}{Input}
\SetKwInOut{Output}{Output}
\SetKwInOut{firststep}{Step 1}
\SetKwInOut{secstep}{Step 2}
\SetKwInOut{thirdstep}{Step 3}
\SetKwInOut{fourthstep}{Step 4}
\DontPrintSemicolon
\Input{Quadrature points and weights
  $\{\bxi^{(i)},\mathtt{w}_i\}_{i=1}^{s_d}$ corresponding to $d$-dimensional, $s_d$-point quadrature
  rule and quadrature points and weights $\{\zeta^{(i)},
  w_i\}_{i=1}^{s_1}$, corresponding to $1$-dimensional,
  $s_1$-point quadrature rule.}
\firststep{Estimate the coefficients $\{f_{\balpha}\}_{\balpha \in
    \calJ_{Q_0}}$ as in eq. (\ref{eq:full_quadr}). }
\secstep{Compute $\bC$ using eq. (\ref{eq:gradient_entries}) and find
  $\bA$ and $\bw$ such that 
\begin{equation*}
\bC = \bA \bLam \bA^T, \ \ \ \bA = [\bw\ \ \bV].
\end{equation*}}
\thirdstep{Estimate $F_{\eta}^{-1}(\cdot)$ and compute $\eta^{(i)} = F_{\eta}^{-1}\left( \frac{\zeta^{(i)}+1}{2}\right)$ and
\begin{equation*}
\hat{\bxi}^{(i)} = \bw \eta^{(i)}, \ \ i = 1,\dots, s_1.
\end{equation*}
}
\fourthstep{Evaluate QoI at $\{\hat{\bxi}^{(i)}\}_{i= 1}^{s_1}$ to obtain
  $\{f(\hat{\bxi}^{(i)})\}_{i=1}^{s_1}$ and estimate the coefficients
  $\{\g_{n}\}_{n=1}^{N_{\zeta}}$ as
\begin{equation*}
\g_{n} \approx \sum_{i=1}^{s_1} f(\hat{\bxi}^{(i)})\psi_{n}(\zeta^{(i)}) w_r.
\end{equation*}}

%\begin{algorithmic}[1]
%\STATE{\textbf{Input:} Quadrature points and weights
%  $\{\bxi^{(i)},\mathtt{w}_i\}_{i=1}^{s_d}$ corresponding to $d$-dimensional, $s_d$-point quadrature
%  rule and quadrature points and weights $\{\zeta^{(i)},
%  w_i\}_{i=1}^{s_1}$, corresponding to $1$-dimensional,
%  $s_1$-point quadrature rule.}
%\STATE{\textbf{Step 1:} Estimate the coefficients $\{f_{\balpha}\}_{\balpha \in
%    \calJ_{Q_0}}$ as in eq. (\ref{eq:full_quadr}). }
%\STATE{\textbf{Step 2:} Compute $\bC$ using eq. (\ref{eq:gradient_entries}) and find
%  $\bA$ and $\bw$ such that 
%\begin{equation*}
%\bC = \bA \bLam \bA^T, \ \ \ \bA = [\bw\ \ \bV].
%\end{equation*}}
%\STATE{\textbf{Step 3:} Estimate $F_{\eta}^{-1}(\cdot)$ and compute $\eta^{(i)} = %F_{\eta}^{-1}\left( \frac{\zeta^{(i)}+1}{2}\right)$ and
%\begin{equation*}
%\hat{\bxi}^{(i)} = \bw \eta^{(i)}, \ \ i = 1,\dots, s_1.
%\end{equation*}
%}
%\STATE{\textbf{Step 4:} Evaluate QoI at $\{\hat{\bxi}^{(i)}\}_{i= 1}^{s_1}$ to obtain
%  $\{f(\hat{\bxi}^{(i)})\}_{i=1}^{s_1}$ and estimate the coefficients
%  $\{\g_{n}\}_{n=1}^{N_{\zeta}}$ as
%\begin{equation*}
%\g_{n} \approx \sum_{i=1}^{s_1} f(\hat{\bxi}^{(i)})\psi_{n}(\zeta^{(i)}) w_r.
%\end{equation*}}

%\Repeat{ change in $\calF_{2}[q]$ is less than a tolerance $\epsilon$}{
 % \For{$i = 1$ to $L$}{
  %  $\{\bmu_i\} \leftarrow \arg\max_{\{\bmu_i\}} \calF_{0}[q]$\\
   % $\{w_i\} \leftarrow \arg\max_{\{w_i\}} \calF_{2}[q]$\\
   % $\{\bSigma_i\} \leftarrow \arg\max_{\left\{\bSigma_i=\operatorname{diag}\left( \sigma_{i1}^2,\dots,\sigma_{id}^2\right)\right\}} \calF_{2}[q]$
  %  }
%}
%\end{algorithmic}
\end{algorithm} 

For the implementation of the above algorithm in the numerical examples that are presented in this paper, the python package \texttt{chaos\char`_basispy} \cite{chaos_basispy2018} has been used that is equipped with polynomial chaos basis function evaluations capabilities, quadrature points generation and computation of the stiffness matrices $\bK_{ij}$ derived above.

\subsection{Error analysis}

The procedure described so far includes computation of the coefficients of a low order polynomial chaos expansion using an efficient numerical integration scheme, typically a low-level sparse quadrature rule. Although the estimation of the coefficients can be satisfactorily accurate, the computation of the gradient vector covariance matrix and subsequently the computation of the active subspace is still subject to the truncation error introduced by using a low order chaos expansion. Below we attempt to quantify this error be deriving some rather standard upper bounds for error of the covariance estimate and its eigenvalues.

Let us denote with $\hat{f}(\bxi)$ the truncated version of $f(\bxi)$ to be used for computation of $\bC$, 
\begin{equation}
\hat{f}(\bxi) = \sum_{\balpha \in \calJ_{Q_0}} f_{\balpha} \psi_{\balpha}(\bxi),
\end{equation}
where $Q_0 < Q$ and write 
\begin{equation}
f(\bxi) = \hat{f}(\bxi) + \epsilon(\bxi).
\end{equation}
The error of approximating the gradient can then be written as 
\begin{equation}
\nabla_{\bxi} \epsilon(\bxi) = \nabla_{\bxi} f(\bxi) - \nabla_{\bxi} \hat{f}(\bxi).
\end{equation}
By defining 
\begin{equation}
\gamma_{Q_0} := ||\nabla_{\bxi} \epsilon(\bxi)||,
\end{equation}
where $||\cdot||$ is the Euclidean norm, we can write 
\begin{equation}
||\nabla_{\bxi} \hat{f}(\bxi) - \nabla_{\bxi} f(\bxi) || \leq \gamma_{Q_0},
\end{equation}
where clearly $\gamma_{Q_0} \to 0$ as $Q_0 \to Q$. Define also
\begin{equation}
\hat{\bC} = \E \left\{\nabla \hat{f}(\bxi) \nabla \hat{f}(\bxi)^T\right\}.
\end{equation} 
By referring to the spectral norm (induced by the Euclidean norm) when $||\cdot||$ is applied on matrices, the following Theorem can be stated:\\
%\begin{theorem} 
%\label{thm:theorem}
\textbf{Theorem 1.} \emph{The norm of the difference between $\bC$ and $\hat{\bC}$ is bounded by 
\begin{displaymath}
||\hat{\bC} - \bC|| \leq \E\left[\gamma_{Q_0}^2\right] + 2\E\left[L\gamma_{Q_0}\right],
\end{displaymath}
where $L = ||\nabla_{\bxi}f(\bxi)||$.
}
%\end{theorem}
%\begin{proof} First note that 
\emph{Proof.} First note that
\begin{eqnarray}\begin{array}{rcl}
||\nabla_{\bxi}\hat{f} + \nabla_{\bxi} f|| & = & ||\nabla_{\bxi} \hat{f} - \nabla_{\bxi} f + 2\nabla_{\bxi} f|| \leq \\ & \leq &  ||\nabla_{\bxi} \hat{f} - \nabla_{\bxi} f|| + 2 ||\nabla_{\bxi}f|| \leq \gamma_{Q_0} + 2L. \end{array}
\end{eqnarray}
Then we have 
\begin{eqnarray}\begin{array}{rcl}
||\nabla_{\bxi} \hat{f} \nabla_{\bxi} \hat{f}^T - \nabla_{\bxi} f \nabla_{\bxi} f^T || & = & \frac{1}{2} ||(\nabla \hat{f} - \nabla f)(\nabla \hat{f} + \nabla f)^T + (\nabla \hat{f} + \nabla f)(\nabla \hat{f} - \nabla f)^T|| \leq \\ & \leq & ||(\nabla \hat{f} + \nabla f)(\nabla \hat{f} - \nabla f)^T|| \leq (\gamma_{Q_0} + 2L) \gamma_{Q_0}. \end{array}
\end{eqnarray}
Finally we get 
\begin{eqnarray}\begin{array}{rcl}
||\hat{\bC} - \bC|| & = & \left|\left|\E \left[\nabla_{\bxi} \hat{f} \nabla_{\bxi} \hat{f}^T\right] - \E\left[\nabla_{\bxi} f \nabla_{\bxi} f^T \right] \right|\right| = \\ & = & \left| \left| \E\left[\nabla_{\bxi} \hat{f} \nabla_{\bxi} \hat{f}^T  - \nabla_{\bxi} f \nabla_{\bxi} f^T \right]\right|\right| \leq \E\left[ \left|\left| \nabla_{\bxi} \hat{f} \nabla_{\bxi} \hat{f}^T  - \nabla_{\bxi} f \nabla_{\bxi} f^T \right|\right|\right] \leq \\ & \leq & \E\left[ \gamma^2_{Q_0}\right] + 2 \E\left[L \gamma_{Q_0}\right] \hspace{7cm} \square
\end{array}
\end{eqnarray} 
%\end{proof}
The following corollary also holds: \\
%\begin{corollary} 
\textbf{Corollary.} \emph{The difference between the $k$th true eigenvalue $\lambda_k$ and the corresponding estimate $\theta_k$ is also bounded as 
\begin{displaymath}
\left|\lambda_k - \theta_k\right| \leq  \E\left[ \gamma^2_{Q_0}\right] + 2 \E\left[L \gamma_{Q_0}\right].
\end{displaymath}
}
%\end{corollary}
%\begin{proof} Simply observe that
\textbf{Proof.} Simply observe that
\begin{displaymath}
\left| \lambda_k - \theta_k\right| \leq \left|\left| \hat{\bC} - \bC \right|\right| \leq  \E\left[ \gamma^2_{Q_0}\right] + 2 \E\left[L \gamma_{Q_0}\right],
\end{displaymath}
where the first inequality follows from Corollary 8.1.6 in \cite{golub} and the second from Theorem 1. $\square$%\cref{thm:theorem}
%\end{proof}

In the above expressions for the error bounds, we can further write 
\begin{equation}
\gamma_{Q_0} = \left\{\sum_{i=1}^d \left(\sum_{|\balpha| =Q_0+1}^Q f_{\balpha} \frac{\partial \psi_{\balpha}(\bxi)}{\partial \xi_i} \right)^2\right\}^{1/2}
\end{equation}
which gives 
\begin{eqnarray}
\E[\gamma_{Q_0}^2] & = & \sum_{i=1}^d \sum_{|\balpha| = Q_0+1}^Q\sum_{|\bbeta| = Q_0+1}^Q f_{\balpha}f_{\bbeta} \E\left[ \frac{\partial \psi_{\balpha}(\bxi)}{\partial \xi_i}\frac{\partial \psi_{\bbeta}(\bxi)}{\partial \xi_i}\right]
\\ & = & \sum_{i=1}^d \sum_{|\balpha|, |\bbeta| = Q_0+1}^Q f_{\balpha} f_{\bbeta} \left(\bK_{ij} \right)_{\balpha\bbeta} \\ & = &
\sum_{i=1}^d \sum_{\substack{|\balpha|, |\bbeta| = Q_0+1 \\ \balpha_{-i} = \bbeta_{-i}}}^Q f_{\balpha}f_{\bbeta} \left(\prod_{\substack{j = 1\\ j\neq i}}^d \E[\psi'_{\alpha_j}(\xi_j)\psi_{\beta_j}(\xi_j)]\right).
\end{eqnarray}
In the above, $\balpha_{-i}$, $\bbeta_{-i}$ indicate the multi-indices $\balpha$, $\bbeta$ where the entries $\alpha_i$, $\beta_i$ respectively, are excluded and the last line follows by using the expressions for $\left(\bK_{ij} \right)_{\balpha\bbeta}$, derived in Appendix \ref{sec:stiff_comp}. For the second term of the error bound, one can simply use the Cauchy-Schwarz inequality to write $\E[L \gamma_{Q_0}] \leq \E[L^2]^{1/2}\E[\gamma_{Q_0}^2]^{1/2}$ and derive a similar expression for $\E[L^2]$, as above. Further numerical investigation of the behavior of the above error bound falls beyond the scope of this work.

\section{Numerical examples}
\label{sec:examples}

\subsection{Polynomial function}

Consider the function $f : \R^d \to \R$ with 
\begin{equation}
\label{eq:quadr_func}
f(\bxi) = a + b \bw^T\bxi + c \bxi^T \bw \bw^T\bxi
\end{equation}
with constants $a$, $b$, $c$ and $\bw \in \R^d$ such that $\bw^T\bw =
1$. For the variables we assume that $\bxi = (\xi_1, \dots, \xi_d)$
with $\xi_i$ independent $\calU(-1,1)$. This gives
\begin{equation}
\nabla f(\bx) = \bw \left(b + 2c \bw^T\bxi\right)
\end{equation}
which gives 
\begin{eqnarray}
\begin{array}{ccl}
\E\left[ \nabla f(\bxi) \nabla f(\bxi)^T \right] & = & \bw \E\left[ (b + 2c
  \bw^T\bxi) (b + 2c \bw^T \bxi ) \right] \bw^T \\ & = &\bw \left(b^2 + 4c^2
  \E\left[ \bxi^T \bw \bw^T\bxi\right] \right) \bw^T \\ & = & 
\bw \left( b^2 + \frac{4}{3}c^2\right) \bw^T
\end{array}
\end{eqnarray}
and it is clear that $b^2 + \frac{4}{3} c^2$ is the only
nonzero eigenvalue corresponding to the eigenvector $\bw$, giving a
1-dimensional active subspace. Setting $\eta = \bw^T \bxi$ we get
\begin{equation}
f(\bxi) = g(\bw^T\bxi)
\end{equation}
where the link function $g(\eta)$ is
\begin{equation}
g(\eta) = a + b \eta + c \eta^2.
\end{equation}

Letting $F_{\eta}$ be the cummulative distribution function of $\eta$,
we are interested in constructing a PC expansion with respect to
$\zeta = (2F_{\eta}(\eta) - 1)$. In our numerical implementation we
use arbitrary values for $a$, $b$, $c$ and $\bw$ that were randomly
generated (and are easily reproducible by fixing the random seed) and we take $d = 10$. We have 
\begin{eqnarray}
\label{eq:w_values}
\begin{array}{r}
\bw = (0.1404, -0.3574, 0.4267, -0.0931, -0.2146, \\ 0.2642,
  0.2560, -0.1895, 0.0046, -0.6680 )^T 
\end{array}
\end{eqnarray}
and 
\begin{eqnarray}
\begin{array}{ccc} a = 1.1500,  & b = 0.9919, & c = 0.9533.
\end{array}
\end{eqnarray}

\begin{figure}[h]
\centerline{
\psfig{figure=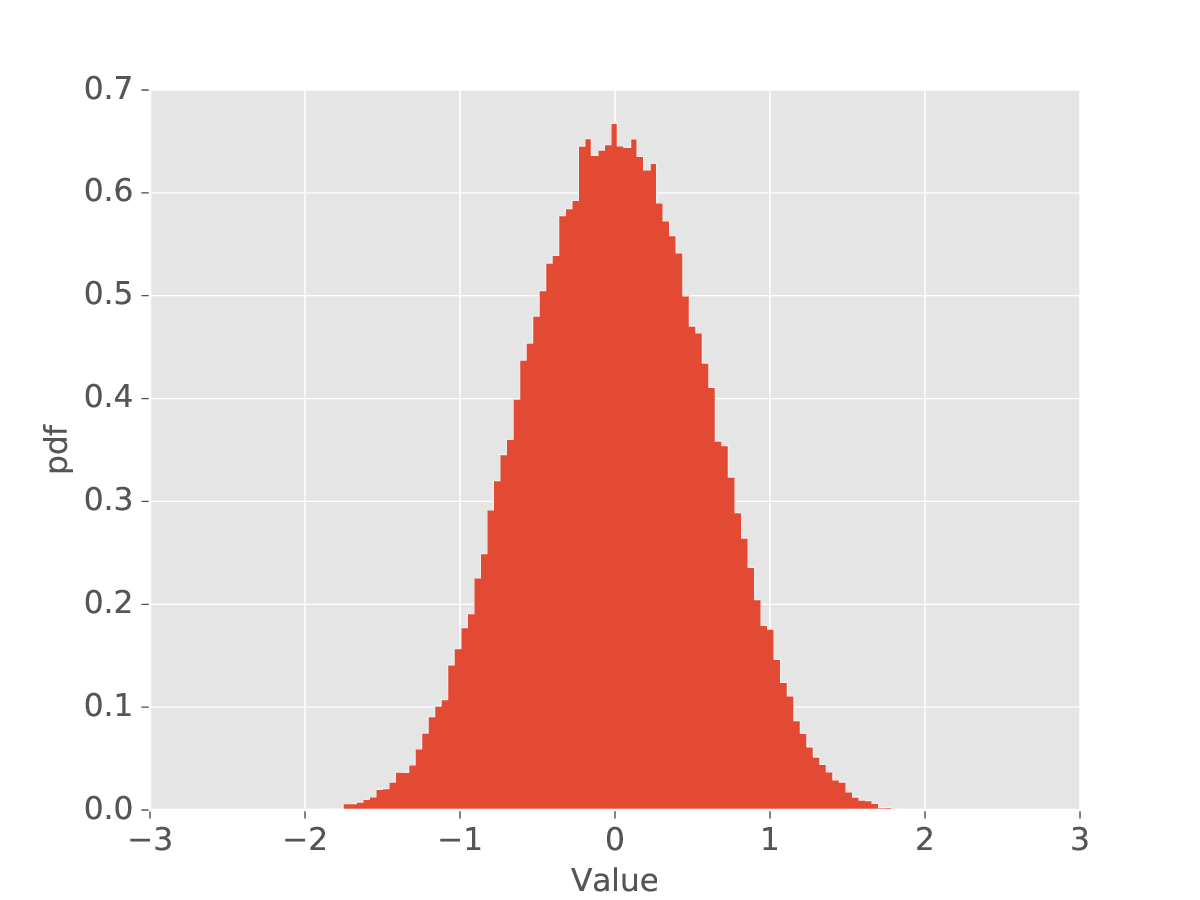,width=0.49\textwidth}
\psfig{figure=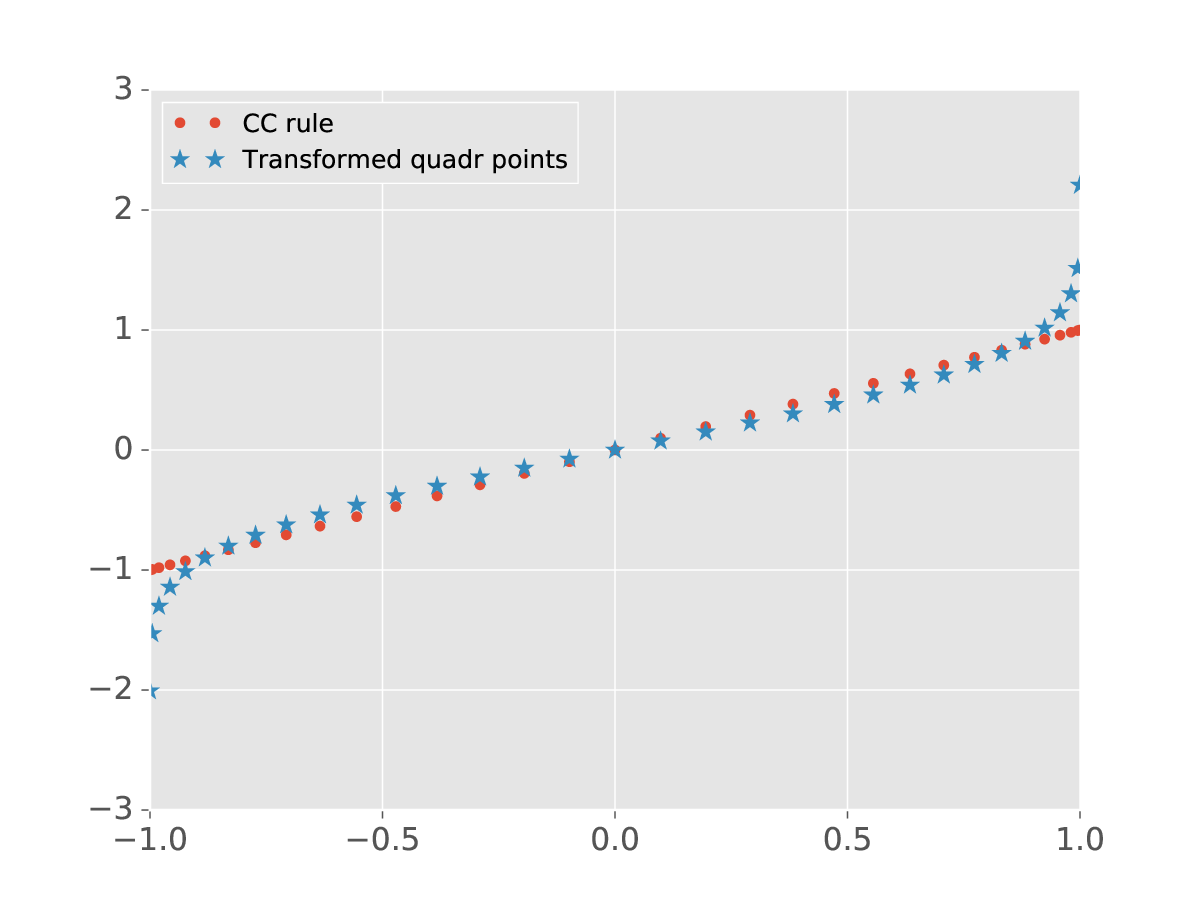, width = 0.49\textwidth}
}
\caption{Left: Plot of the empirical pdf of $\eta = \bw^T \bxi$ for $\bw$
  given in eq. (\ref{eq:w_values}). Right: Level-5 Clenshaw-Curtis quadrature
  points on $[-1,1]$ and their mappings on $\eta$-space.\label{fig:pdf_eta}}
\end{figure}

\begin{figure}[h]
\centerline{
	\psfig{figure=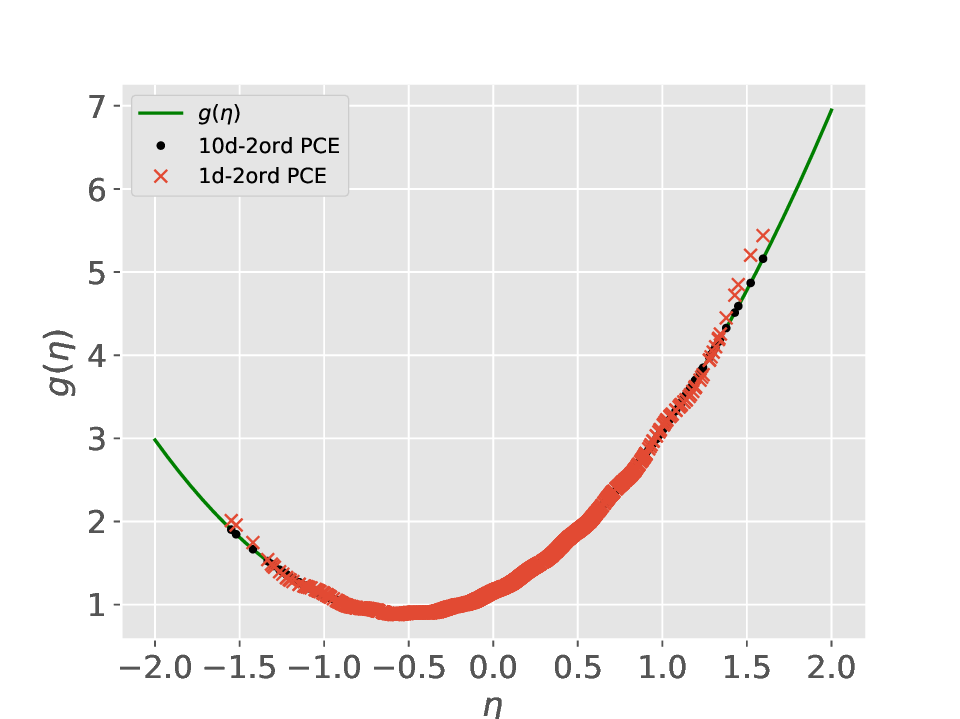, width = 0.49\textwidth}
	\psfig{figure=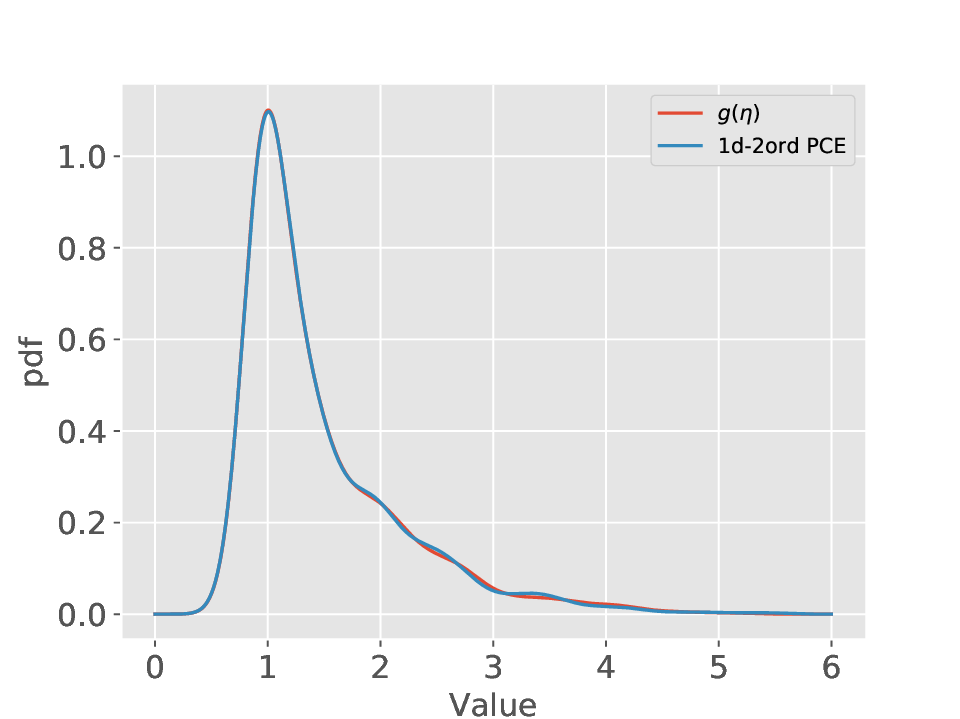, width = 0.49\textwidth}
}
\caption{Left: Predictions of the link function using 1d- and 10d- expansions. Right: Comparison of the true and the estimated density function bases on 1d-20th-order chaos expansion.}
\label{fig:quadr_func}
\end{figure}

Fig. \ref{fig:pdf_eta} (left) shows the empirical pdf of $\eta$ for the above
choice of $\bw$ and Fig. \ref{fig:pdf_eta} (right) shows the quadrature
points of a level-5 Clenshaw-Curtis rule and their transformed values on
the $\eta$-space through the mapping $F_{\eta}^{-1}(\frac{\zeta +
  1}{2})$. Note here that the expression (\ref{eq:quadr_func}) can be
rearranged as a series of Legendre polynomials, therefore its PC
expansion is essentially known and has exact order 2. The same expansion can
be computed numerically using sparse quadrature rule and specifically,
a level-2 rule suffices for an accurate estimation, corresponding to $221$
evaluations. Estimation of the coefficients of a 1d PC expansion with
respect to $\zeta$ using the known orthogonal transformation $\bw$ can
be achieved using a level-5 rule that corresponds to $33$ evaluations
of $f$. Note that the PC expansion with respect to $\zeta$ is no
longer a $2$nd order series. The order of polynomials necessary to
achieve convergence here depend on the nature of the inverse cdf
$F_{\eta}^{-1}$. In our case we find that a $20$th
 order polynomial chaos expansion suffices to achieve a low
 error. Fig. \ref{fig:quadr_func} shows the resulting
 expansions evaluated at $1000$ sample points $\bxi$ or their
 corresponding $\zeta = 2F_{\eta}(\bw^T\bxi) - 1$ accordingly and the
 density function of the output. Note
 that in order to obtain a common input
 domain the plots show the dependence of the expansions with
 respect to the $\eta$ sample points.

%\begin{figure}[t]
%\centering
%\subfloat[Empirical pdf of $\eta$.]{\label{fig:pdf_eta}\includegraphics[width = 0.5\textwidth]{./images/quadr_func_eta_pdf.eps}}
%\subfloat[Quadrature points in $\zeta$- and $\eta$-space.]{\label{fig:quadr_points}\includegraphics[width = 0.5\textwidth]{./images/quadr_func_eta_quadr_rule.eps}}
%\caption{Left: Plot of the empirical pdf of $\eta = \bw^T \bxi$ for $\bw$
%  given in eq. (\ref{eq:w_values}). Right: Level-5 Clenshaw-Curtis quadrature
%  points on $[-1,1]$ and their mappings on $\eta$-space.}
%\end{figure}

%\begin{figure}[t]
%\centering
%\subfloat[MC function output sample comparison.]{\label{fig:quadr_func}\includegraphics[width = 0.49\textwidth]{./images/quadr_func_pces.eps}}
%\subfloat[Pdf comparison]{\label{fig:quadr_func_pdf}\includegraphics[width = 0.49\textwidth]{./images/quadr_func_pdf.eps}}
%\caption{Left: Predictions of the link function using 1d- and 10d- expansions. Right: Comparison of the true and the estimated density function bases on 1d-20th-order chaos expansion.}
%\end{figure}

\subsection{Multiphase flow: 1d transport and decay of ammonium}

Multiphase-multicomponent flow and transport models
are known for their particular challenges associated with the
uncertainty propagation problem which are mainly due to the highly
nonlinear nature of the coupled systems of
equations that describe the complex physical process and the varying
role of the many random parameters involved in the system. Here we
consider the problem of a 1-dimensional transport of the microbially
mediated first-order oxidation of ammonium (NH$^+_4$) to nitrite (NO$^-_2$) and the
subsequent oxidation of nitrite to its ion (NO$^-_3$) that has been
previously investigated in \cite{cho_1971, mcnab, vangenuchten}. The flow domain under consideration where the flow of ammonium takes places
is taken here to be a one-dimensional column of 2 meters length. 

%\subsubsection{The model}

%To
%solve for the mass fractions of the species under investigation, 
For the numerical solution of the transport and decay model, we employ the
multiphase-multicomponent simulator TOUGH2 \cite{pruess} that solves
the integral form of the system of governing equations using the integral
finite difference method. More specifically, the EOS7r module \cite{olden} that specializes in modeling
radionuclide transport, is used here, allowing the description of the
oxidation effect in a similar manner as specifying
half-life parameters to describe radioactive or
biodegradation effects. Detailed description of the balance equations and capillary pressure model can be found in \cite{pruess}. The mass components considered here are water, air, ammonium and nitrite and the fluid phases are aqueous and gas. We consider a first-order decay law for the mass components \cite{olden}.

%The implementation of the first-order decay
%effect described above is performed by considering that the source
%term accounts for the decay and we write the mass accumulation condition
%for $k = am$, $n$ as
%\begin{equation}
%\displaystyle{\frac{dM^k}{dt} = \nabla \cdot \bF^k - \lambda_k M^k}.
%\end{equation}

\begin{figure}
\centering
\includegraphics[width = 0.8\textwidth]{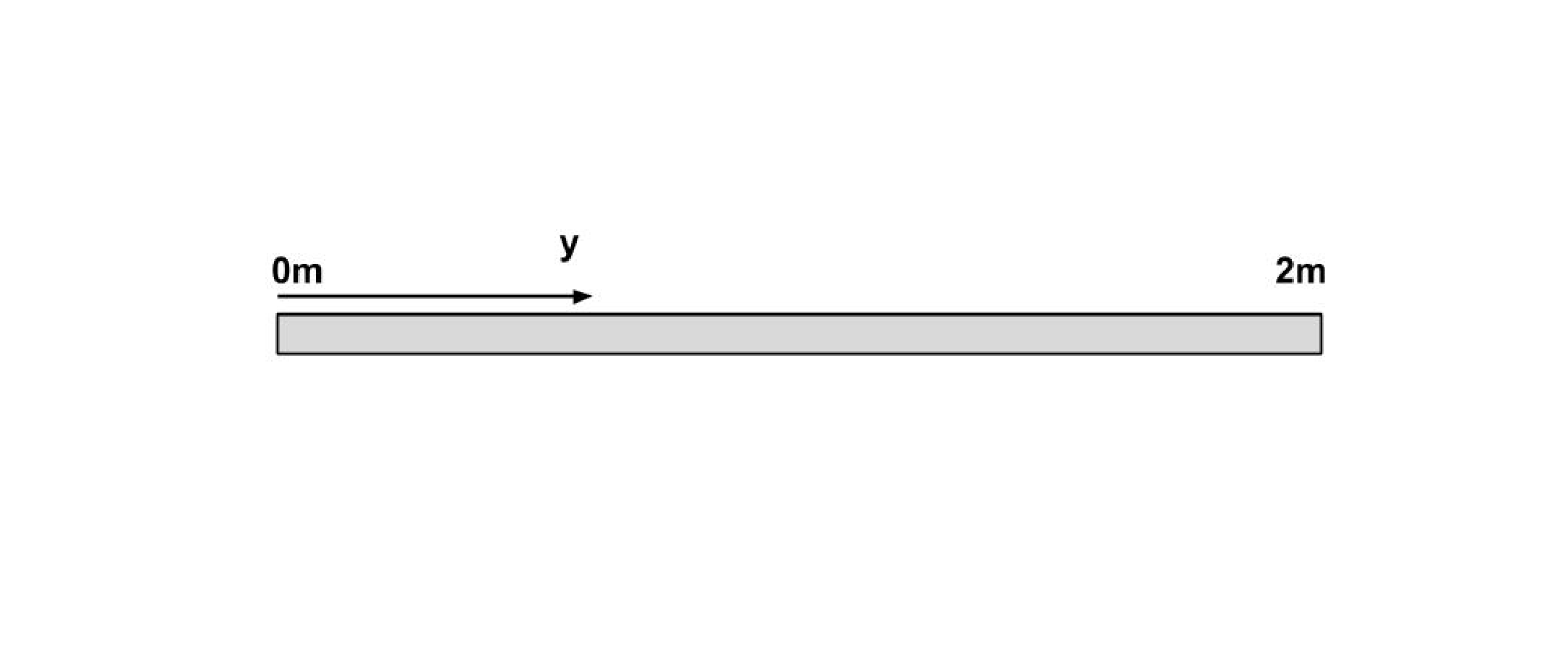}
\caption{Domain of the ammonium transport problem. Ammonium is
  injected at $y = 0$. \label{fig:domain}}
\end{figure}

In the scenario considered here, the domain is discretized into $400$ blocks, each
of length $0.005$m. and a source of ammonium is placed on the one
boundary ($y = 0$), while no concentration of nitrite is initially
present. The domain is shown in Fig. \ref{fig:domain}. Uncertainty is introduced in the
model through six model parameters, namely the decay parameters of
ammonium and nitrite, $T_{1/2}^{am}$ and $T_{1/2}^n$ respectively, the
constant flux of ammonium at the boundary or more precisely the
constant injected concentration  $\chi_0^{am}$, the
distribution coefficient $Kd_w^{am}$ of ammonium that affects its adsorption onto
the immobile solid grains and the porosity $\phi$ of the medium. All
other model parameters were assigned fixed values which are shown in
Table \ref{tab:tough2_params}. All random input parameters, denoted in a
vector form as $\btheta = (T_{1/2}^{am}, T_{1/2}^{am}, \chi_0^{am},
Kd_w^{am}, \phi)^T$ are assumed to
follow uniform distributions $\theta_i \sim \calU(\theta_l^i,
\theta_u^i)$ where the intervals $[\theta_l^i,
\theta_u^i]$, $i = 1, \dots, 5$ are given in Table
\ref{tab:priors}. Next, for the construction of a polynomial chaos
approximation of model output, the random input parameters are
rescaled to $\calU(-1, 1)$ random variables via the transformation
\begin{equation}
\xi_i = 2\frac{\theta^i - \theta_l^i}{\theta^i_u - \theta^i_l} - 1.
\end{equation}

\begin{table}[h]
\caption{Material and initial parameters used in my simulations.}
\label{tab:tough2_params}
\centering
\begin{tabular}{l | c | c}
Parameter & Symbol & Value \\
\hline
Rock grain density & $\rho_R$ & $2650$kg/m$^3$ \\
Tortuosity factor & $\tau_0$ & $1$ \\
Absolute permeability & $\kappa$ & $10^{-12}$m$^2$ \\
Diffusivities (all $k$): gas phase& $ d_{g}^k $ & $10^{-6}$ \\
aqueous phase & $d_{l}^k$ & $10^{-10}$\\
Molecular weight & - & $18$ \\
Inverse Herny's constant & - & $10^{30}$ \\
Initial pressure & $P(0)$ & $1.010\cdot 10^{5}$ \\
Initial gas saturation & $S_g$ & $0.75$ \\
Temperature (constant) & $T$ & $25^{\circ}$ \\
\end{tabular}
\end{table}

\begin{table}[h]
\caption{Input parameter value range}
\label{tab:priors}
\centering
\begin{tabular}{l|c|c}
Parameter & $\theta_l$ & $\theta_u$ \\
\hline
$T_{1/2}^{am}$ & $10^5$ & $10^6$\\
$T_{1/2}^{n}$ & $10^4$ & $5\cdot 10^5$ \\
$\chi_0^{am}$ & $0.008$ & $0.015$ \\
$Kd_w^{am}$ & $9\cdot 10^{-5}$ & $1.1\cdot 10^{-4}$ \\
$\phi$ & $0.2$ & $0.6$ 
\end{tabular}
\end{table}

\subsubsection{Results}

\begin{figure}[t]
\centerline{
	\psfig{figure=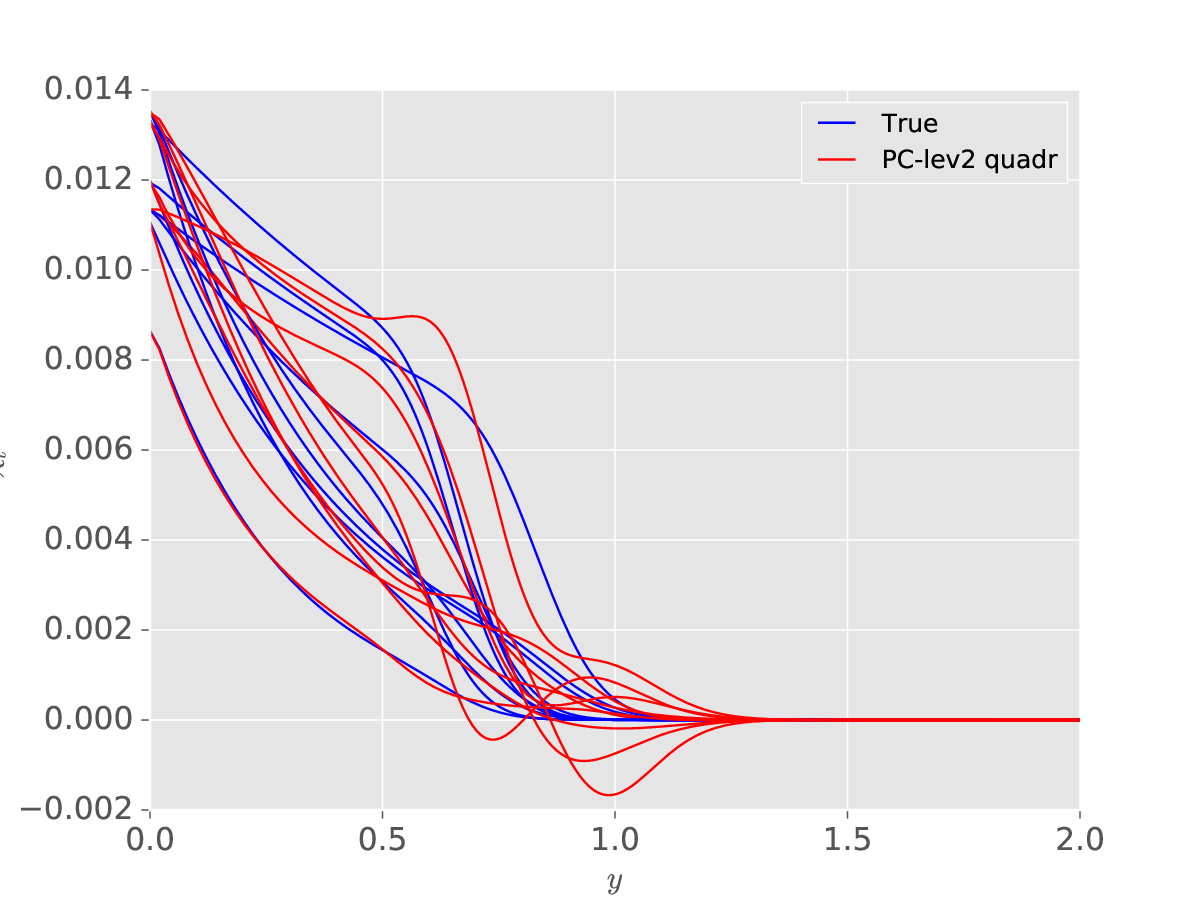, width= 0.49\textwidth}
	\psfig{figure=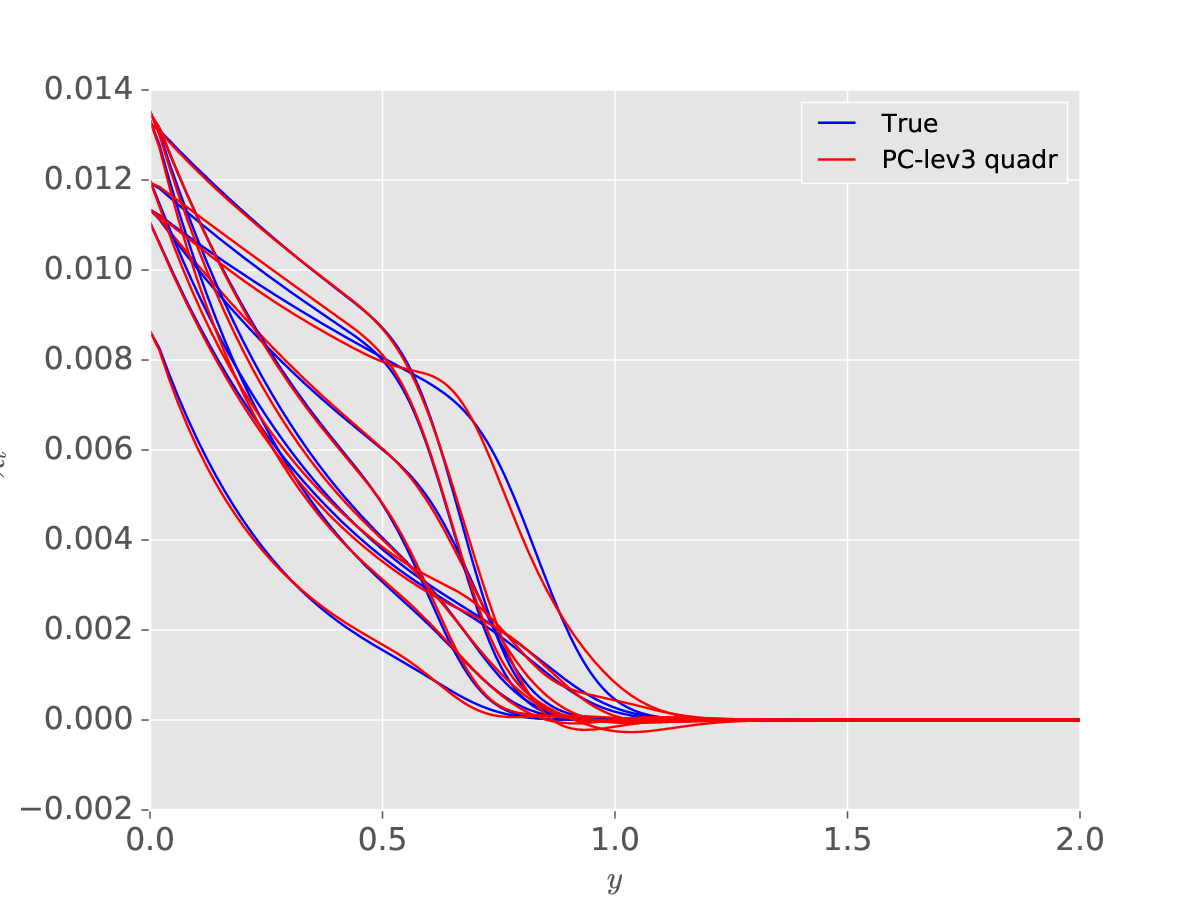, width= 0.49\textwidth}
}
\centerline{
	\psfig{figure=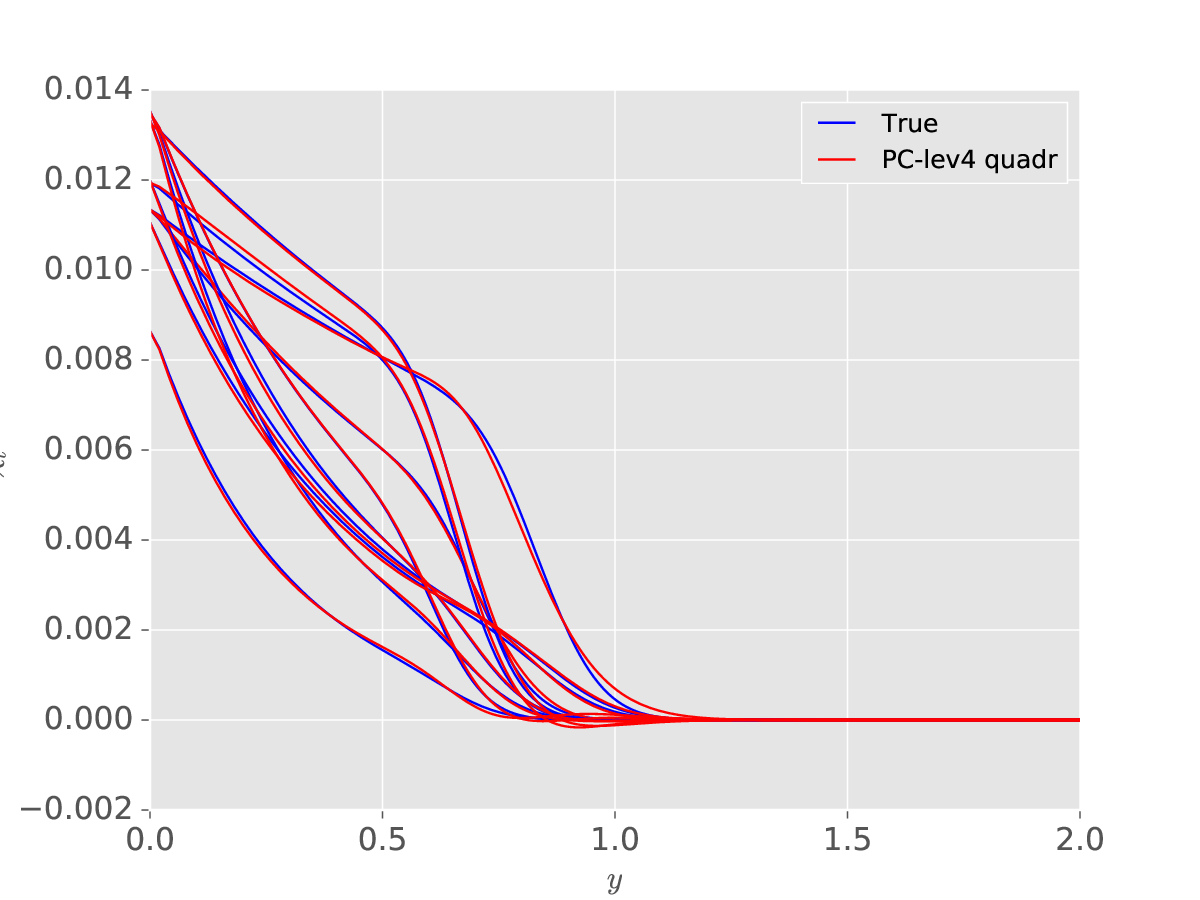, width= 0.49\textwidth}
	\psfig{figure=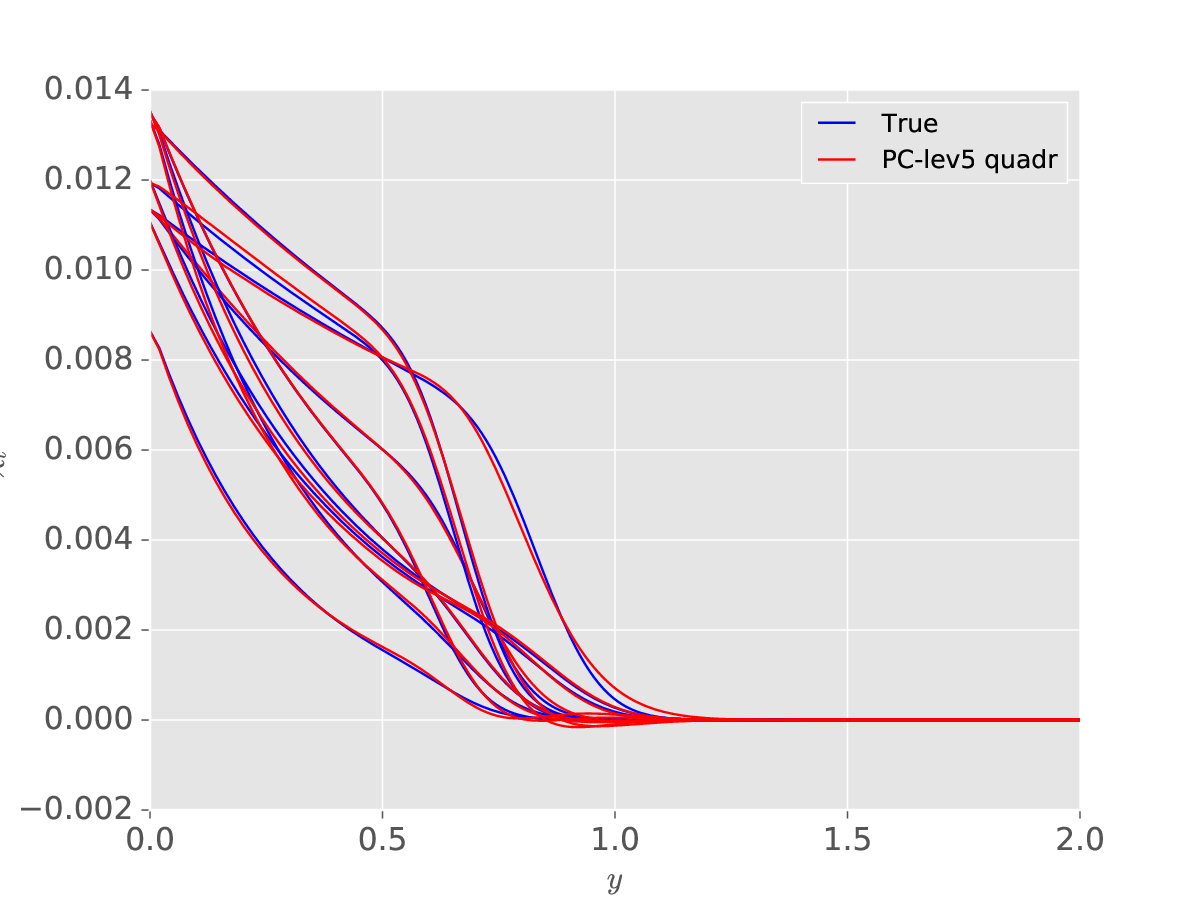, width= 0.49\textwidth}
}
\caption{$10$ Monte Carlo realizations of the model outputs and their
  corresponding PC approximations using quadrature level $\ell = 2$,
  $3$, $4$ and $5$ respectively.\label{fig:full_pces}}
\end{figure}

\begin{figure}
\centerline{
	\psfig{figure=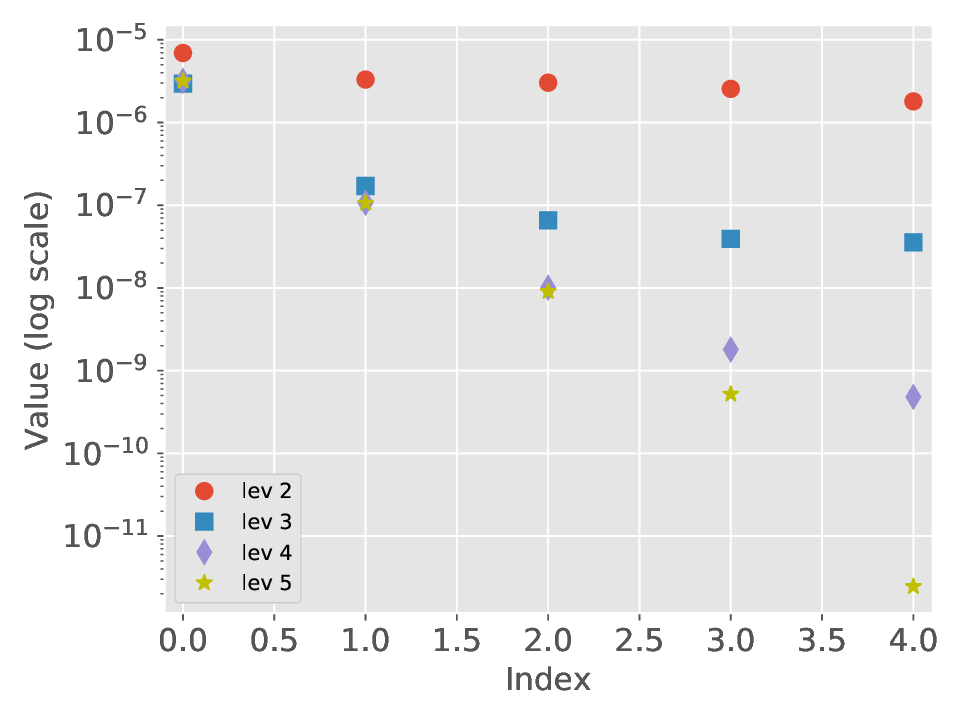, width=0.49\textwidth}
	\psfig{figure=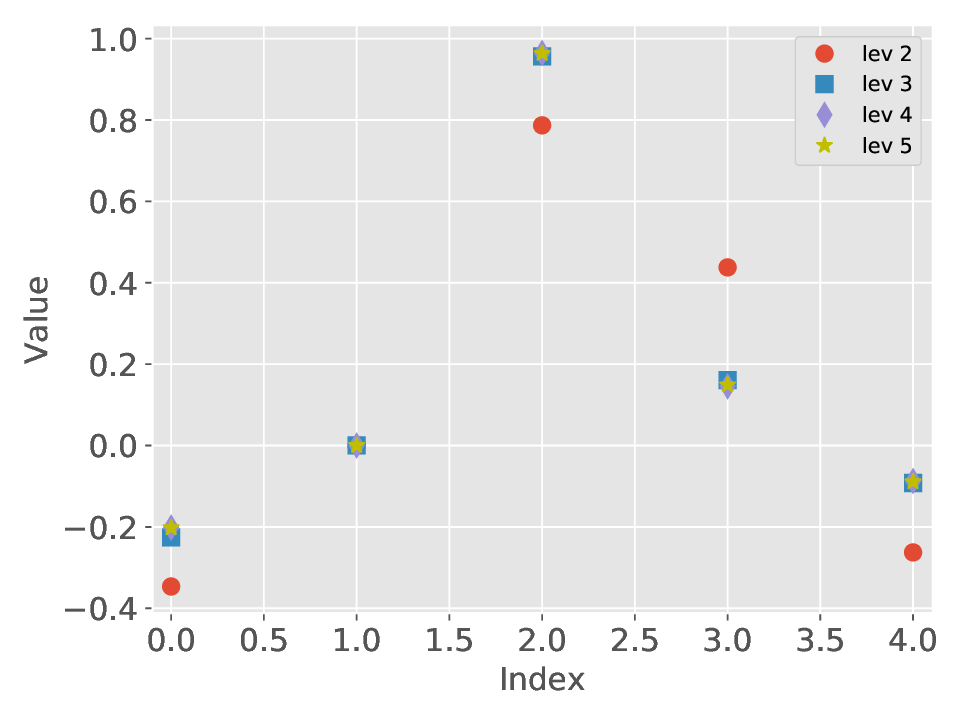, width=0.49\textwidth}
}
\caption{Eigenvalues of the gradient matrix (left) and the values of
  the dominant eigenvector $\bw$ (right) for chaos coefficients
  corresponding to different quadrature levels. \label{fig:eigenpairs}}
\end{figure}

\begin{figure}[h]
\centerline{
	\psfig{figure=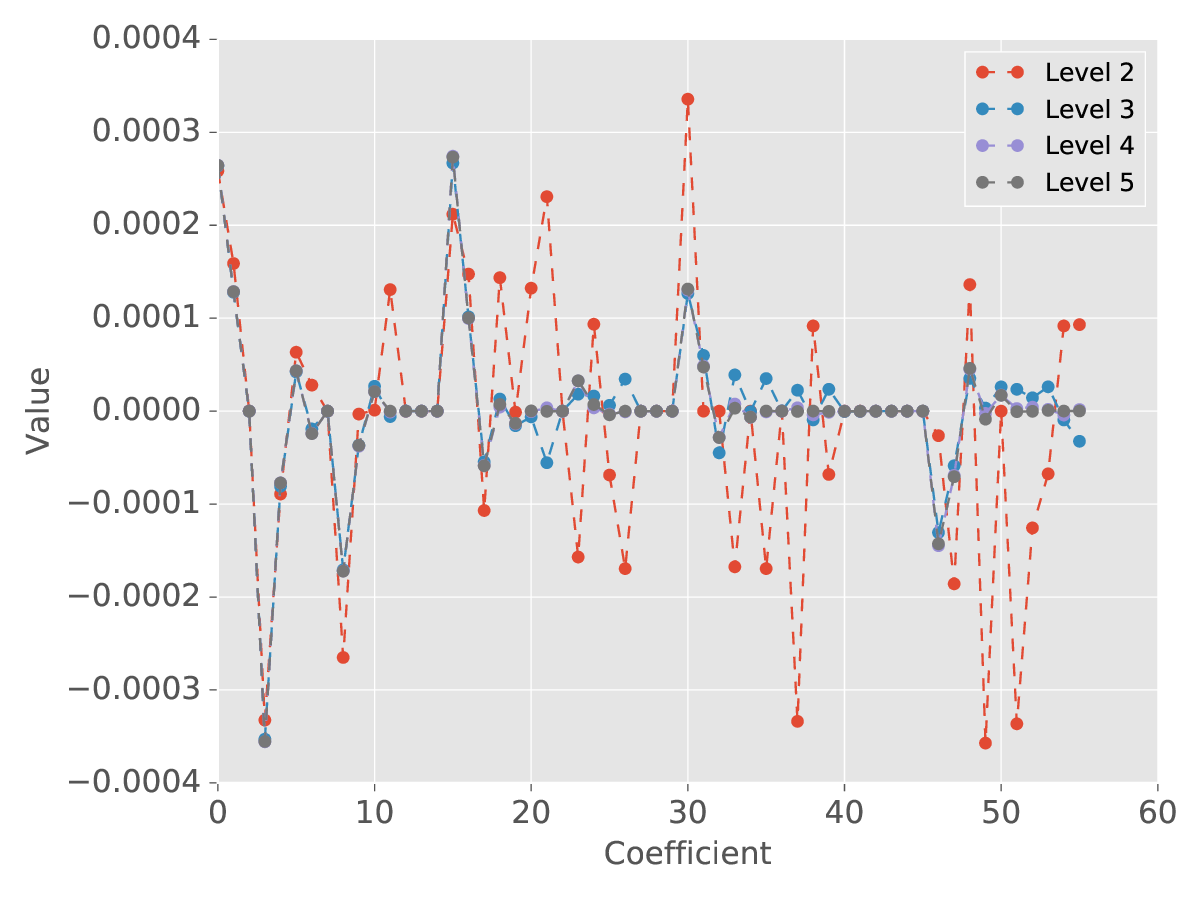, width = 0.49\textwidth}
	\psfig{figure=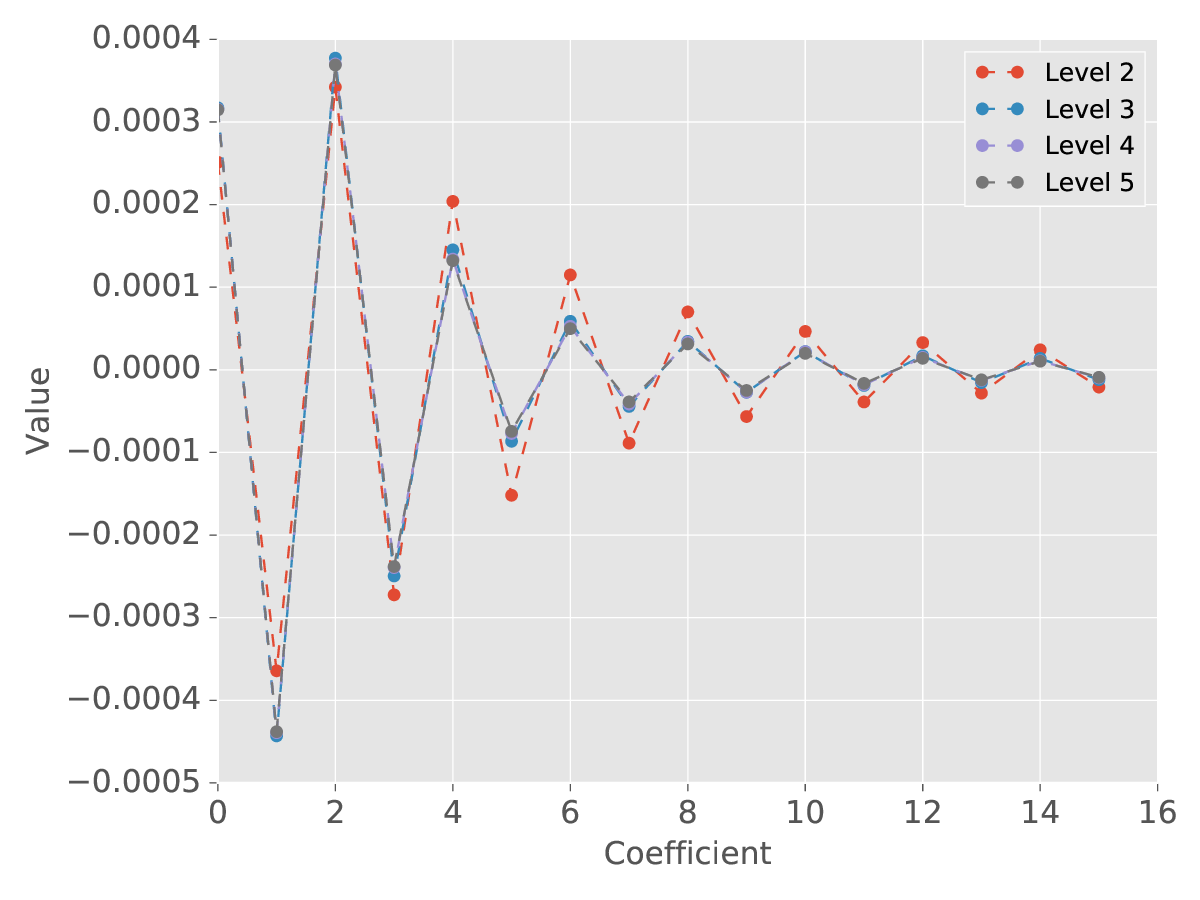, width = 0.49\textwidth}
}
\caption{Estimated coefficients of the chaos expansions using different quadrature rule levels. \label{fig:coeffs_5d}}
\end{figure}

The transport model runs for a time period corresponding to $t = 200$
hours and the mass fraction of ammonium in the aqueous phase is observed. The quantity of
interest considered here is the mass fraction $\chi_t^{am}(y)$ at the
middle point of the column ($y = 1$m distance from the source), where
neglect the index $w$ for simplicity. First, to illustrate the
motivation for applying a basis reduction procedure on the QoI, we
compute the polynomial chaos expansion of the mass fraction of
ammonium over the whole domain $y \in [0,2]$. Figure
\ref{fig:full_pces} shows a comparison of $10$ Monte Carlo outputs of
the model with the polynomials chaos expansions evaluated at the same
MC inputs. The four figures correspond to the cases where the
coefficients have been estimated using a Clenshaw-Curtis quadrature
rule of level 2, 3, 4 and 5 respectively. Qualitatively we observe
that for level 3 and higher, the surrogate is a very good
approximation of the model. However, at $y = 1$ we can see that the
chaos expansions of $\chi_t^{am}(y)$ fall sometimes below zero which
is not an acceptable value since $\chi_t(y)^{am} \geq 0$. This
immediately makes the chaos expansiom inaccurate not only as a
surrogate for fast computation of the model output but will also result in
false empirical distributions of the QoI. 
Next, we use the estimated coefficients for estimation of the gradient
matrix $G$ and we denote with $G^\ell$ the gradient matrix estimated
using the coefficients from level $\ell$ quadrature rule. The
eigenvalues and the dominating eigenvector $\bw^{\ell}$ are shown in
Fig. \ref{fig:eigenpairs}. For the cases $\ell = 3$, $4$ and $5$
the eigenvalues and the vector $\bw^{\ell}$ almost coincide, indicating
convergence of the eigenvalues and existence of a 1-dimensional active
subspace. In the $\ell = 2$ case, the eigenvalues seem to be inaccurate and
perhaps the projection vector does not define an optimal
rotation for dimension reduction. In addition, even though a ``gap"
can be seen between the first and the remaining eigenvalues, their
values indicate that they are not negligible. Fig. \ref{fig:coeffs_5d} (left) shows the $56$ estimated coefficients of the expansions for the different quadrature rule levels and similar conclusions can be drawn regarding the convergence of the numerical integration and the accuracy of the expansion, as the level-2 results are completely ``off" comparing to the remaining cases and even several coefficients of the level-3 case display some discrepancy before being stabilized in level-4 and level-5 rules. 

%\begin{figure}[t]
%\centering
%\includegraphics[width = 0.49\textwidth]{./images/full_pce_lev2.eps}
%\includegraphics[width = 0.49\textwidth]{./images/full_pce_lev3.eps}
%\includegraphics[width = 0.49\textwidth]{./images/full_pce_lev4.eps}
%\includegraphics[width = 0.49\textwidth]{./images/full_pce_lev5.eps}
%\caption{$10$ Monte Carlo realizations of the model outputs and their
%  corresponding PC approximations using quadrature level $\ell = 2$,
%  $3$, $4$ and $5$ respectively.\label{fig:full_pces}}
%\end{figure}

%\begin{figure}[h]
%\centering
%\subfloat[Eigenvalues]{
%\includegraphics[width = 0.49\textwidth]{./images/eigenvals.eps}
%\subfloat[Dominant eigenvector]{
%\includegraphics[width = 0.49\textwidth]{./images/eigenpairs.eps}
%\caption{Eigenvalues of the gradient matrix (left) and the values of
%  the dominant eigenvector $\bw$ (right) for chaos coefficients
%  corresponding to different quadrature levels. \label{fig:eigenpairs}}
%\end{figure}

%\begin{figure}[h]
%\centering
%\subfloat[5d expansions]{\label{fig:coeffs_5d}\includegraphics[width = 0.49\textwidth]{./images/ammon_pce_coeffs_5d.eps}}
%\subfloat[1d expansions]{\label{fig:coeffs_1d}\includegraphics[width = 0.49\textwidth]{./images/ammon_pce_coeffs_1d.eps}}
%\caption{Estimated coefficients of the chaos expansions using different quadrature rule levels.}
%\end{figure}

\begin{figure}
\centerline{
	\psfig{figure=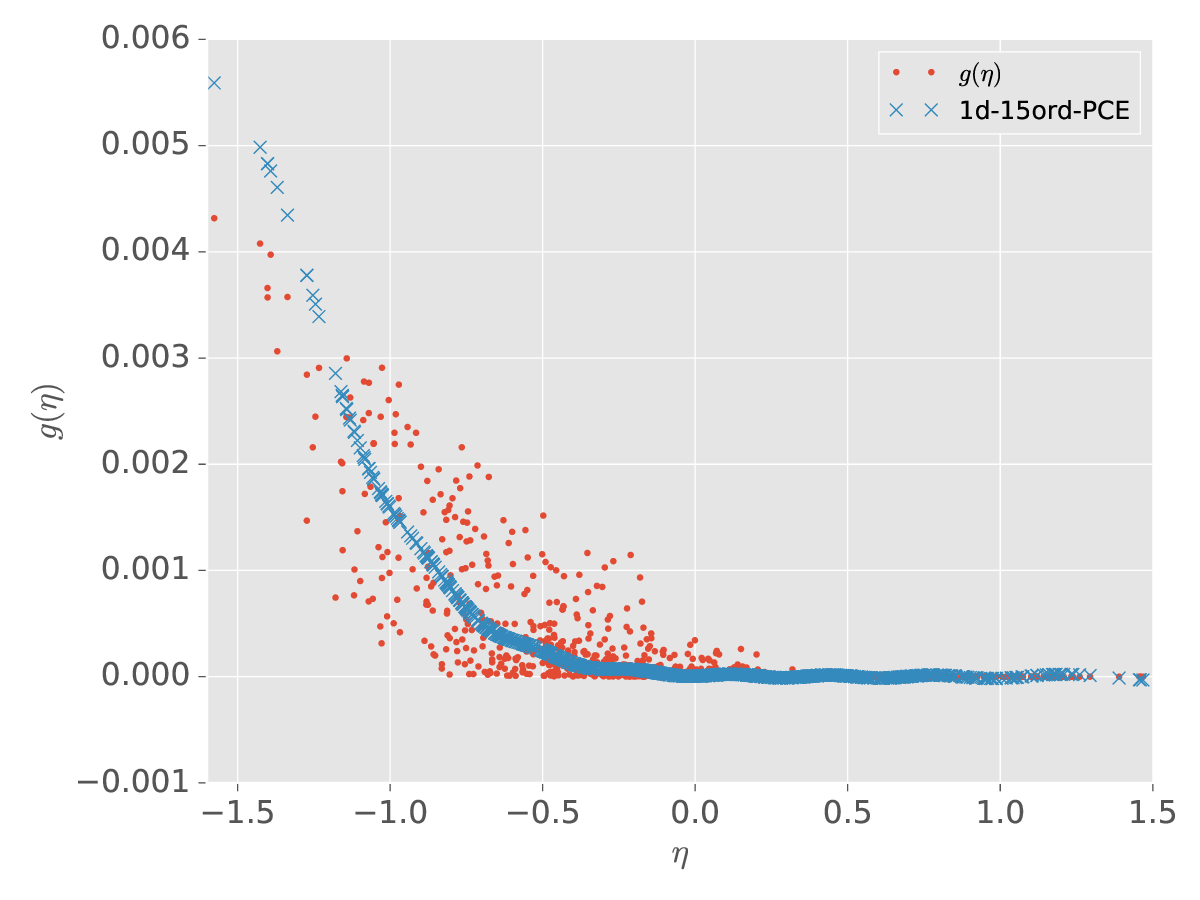, width = 0.49\textwidth}
	\psfig{figure=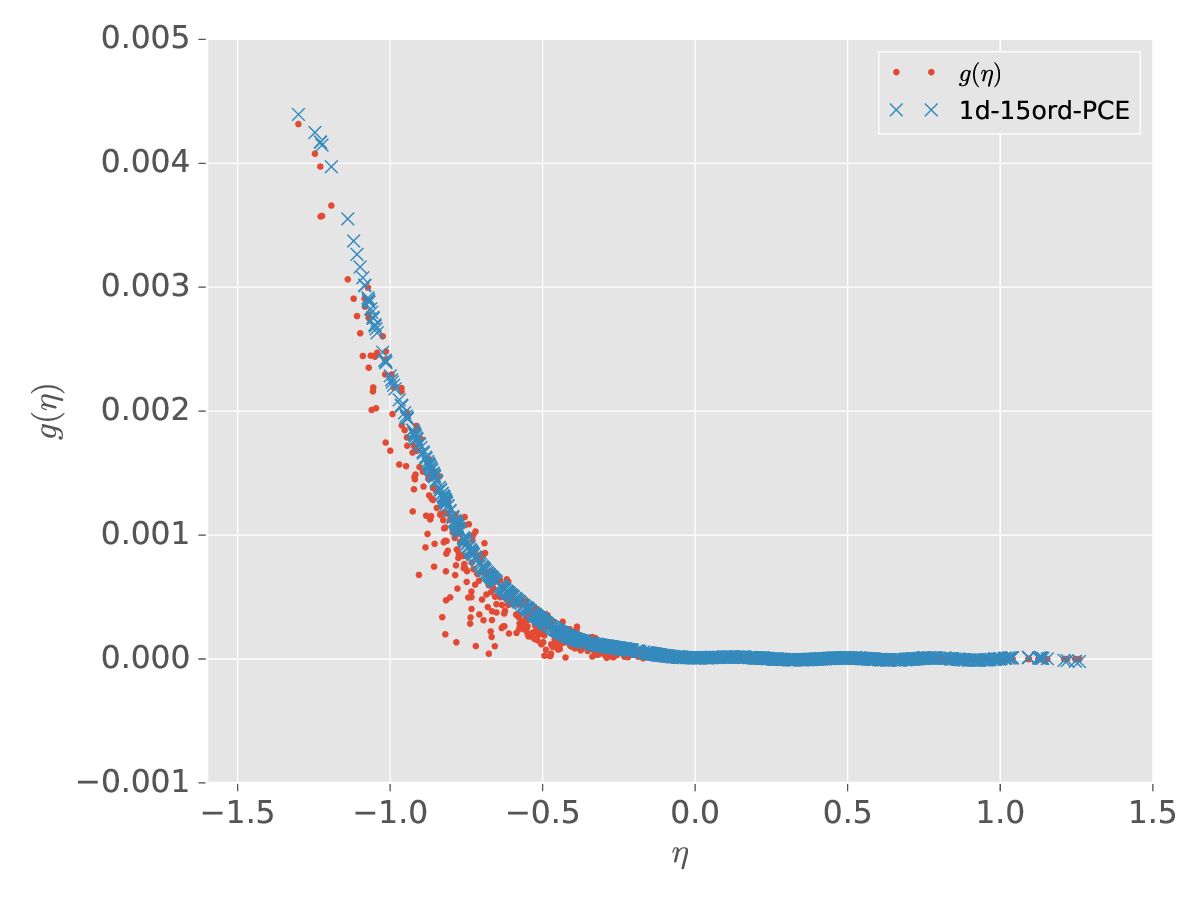, width = 0.49\textwidth}
}
\centerline{
	\psfig{figure=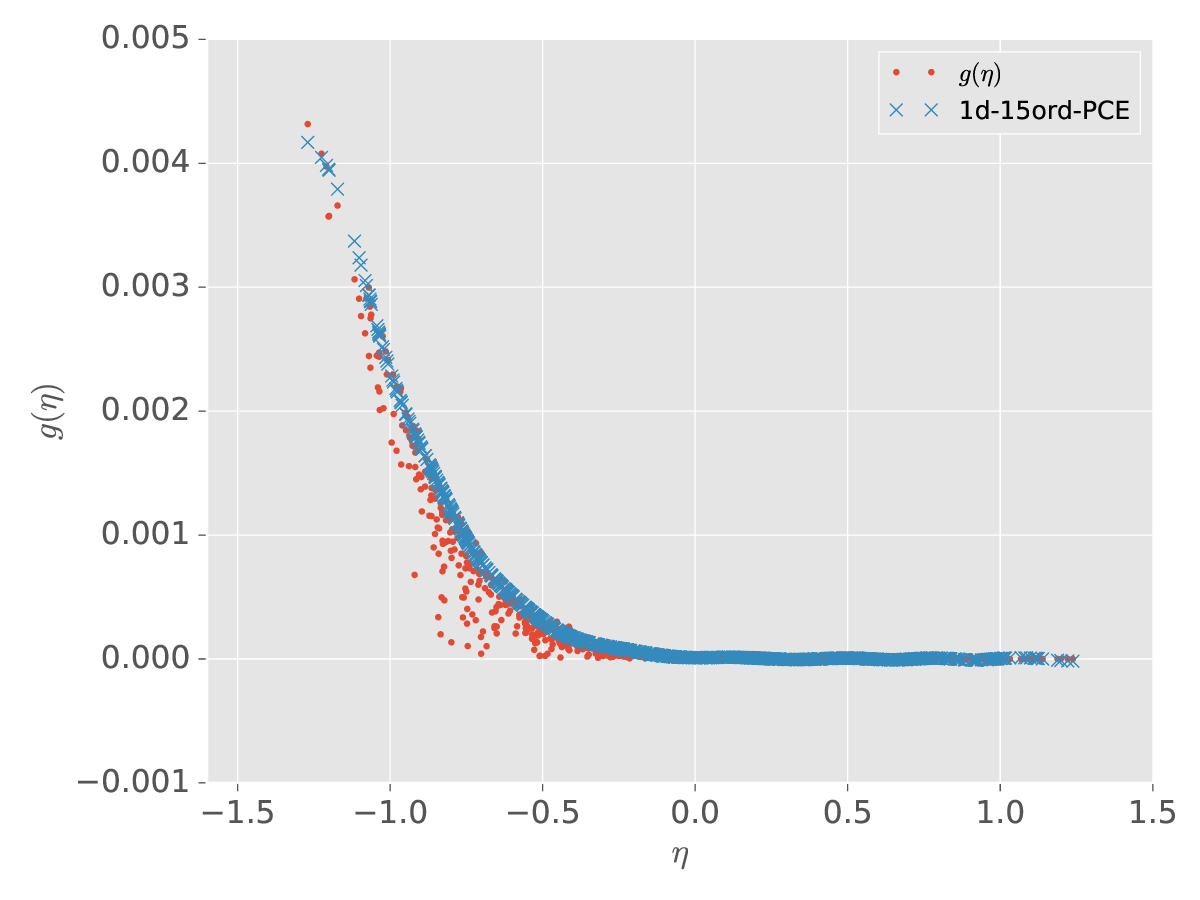, width = 0.49\textwidth}
	\psfig{figure=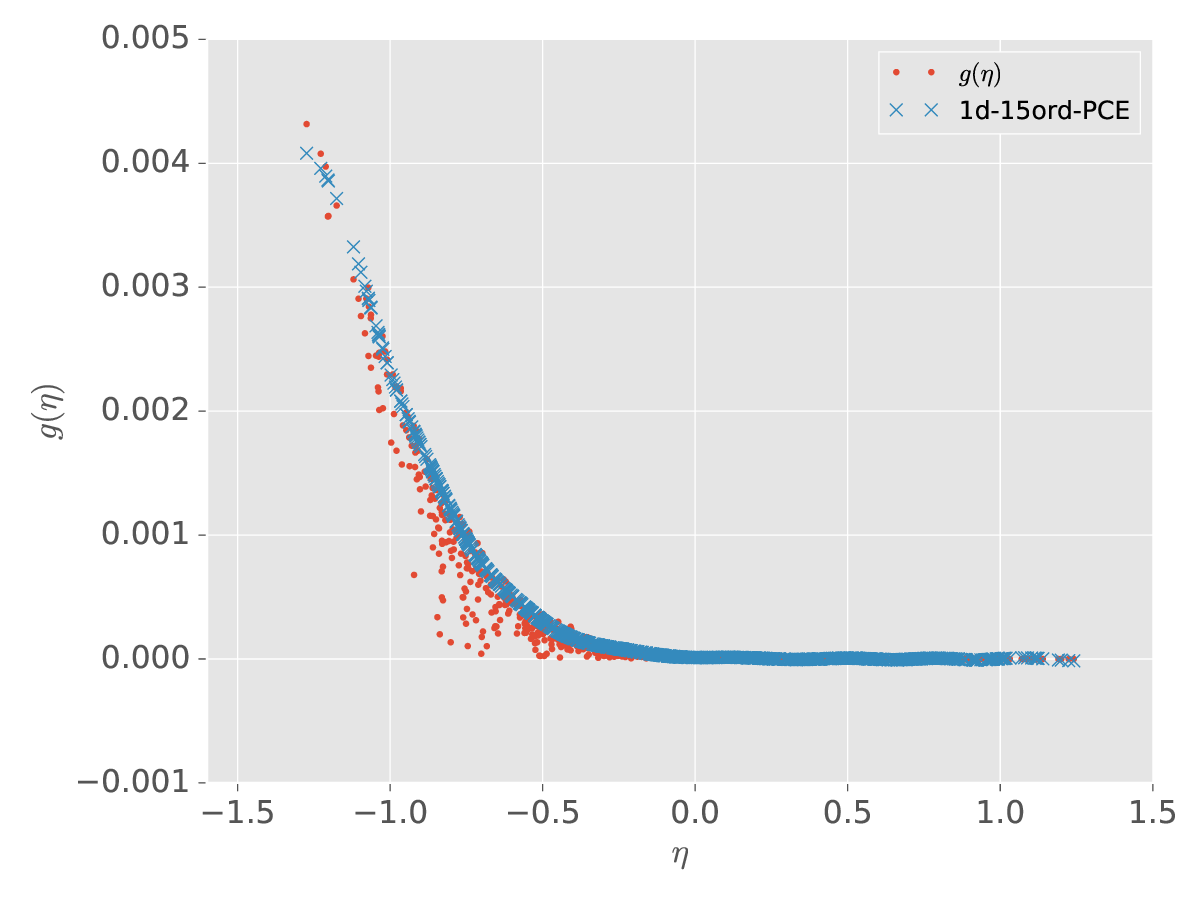, width = 0.49\textwidth}
}
\caption{Evaluation of the model output and the 1-d PC expansions on
  $1000$ MC samples of $\eta^{\ell}$ for $\ell = 2$, $3$, $4$ and
  $5$. \label{fig:1d_pces_comparison}}
\end{figure}

After defining the random \emph{active} variables $\eta^{\ell} = \bw^{\ell T} \bxi$ with
cdf $F_{\eta^{\ell}}(\cdot)$ corresponding to the active subspaces revealed by the chaos expansions estimated above, we construct the PC approximations with respect
to the germs 
\begin{equation}
\zeta^{\ell} = 2F_{\eta^{\ell}}(\eta^{\ell}) - 1, \ \ \ell = 2,\dots, 5,
\end{equation}
of order $15$ using the level-$5$ 1-dimensional CC quadrature
rule. The estimated coefficients are displayed in Fig. \ref{fig:coeffs_5d} (right), where except the one corresponding the level 2 rule, they seem to be in good aggrement, as expected from the similarity of their respective projection vectors $\bw^{\ell}$. Fig. \ref{fig:1d_pces_comparison} shows values of $1000$ MC
outputs of the true model and the 1-d PC expansion as a
function the common input $\eta^{\ell}$. Evaluation of the full
model is carried out at the points $\bxi^{(i)} = \bw^{\ell}
\eta^{\ell}$$^{(i)}$ while evaluation of the 1-d PC expansions is
obtained at $\zeta^{(i)} = F_{\eta^{\ell}}(\eta^{\ell}$$^{(i)})$. It is
observed again that for $\ell \geq 3$, the one dimensional
representation of $\chi_t^{am}$ provides a quite accurate
approximation of the true QoI whose scatter around the 1-d curve due
to the influence of the othogonal subspace $\rm span$$\left\{\bw^T\bxi \right\}^T$ is
very small.

%\begin{figure}
%\centering
%\subfloat[Level 2]{\includegraphics[width = 0.5\textwidth]{./images/ammon_func_pces_lev2.eps}}
%\caption{Level 2}
%\end{subfigure}
%\subfloat[Level 3]{\includegraphics[width = 0.5\textwidth]{./images/ammon_func_pces_lev3.eps}}\\
%\caption{Level 3}
%\end{subfigure}
%\subfloat[Level 4]{\includegraphics[width = 0.5\textwidth]{./images/ammon_func_pces_lev4.eps}}
%\caption{Level 4}
%\end{subfigure}
%\subfloat[Level 5]{\includegraphics[width = 0.5\textwidth]{./images/ammon_func_pces_lev5.eps}}
%\caption{Level 5}
%\end{subfigure}
%\caption{Evaluation of the model output and the 1-d PC expansions on
%  $1000$ MC samples of $\eta^{\ell}$ for $\ell = 2$, $3$, $4$ and
%  $5$. \label{fig:1d_pces_comparison}}
%\end{figure}

\begin{figure}[h]
\centerline{
	\psfig{figure=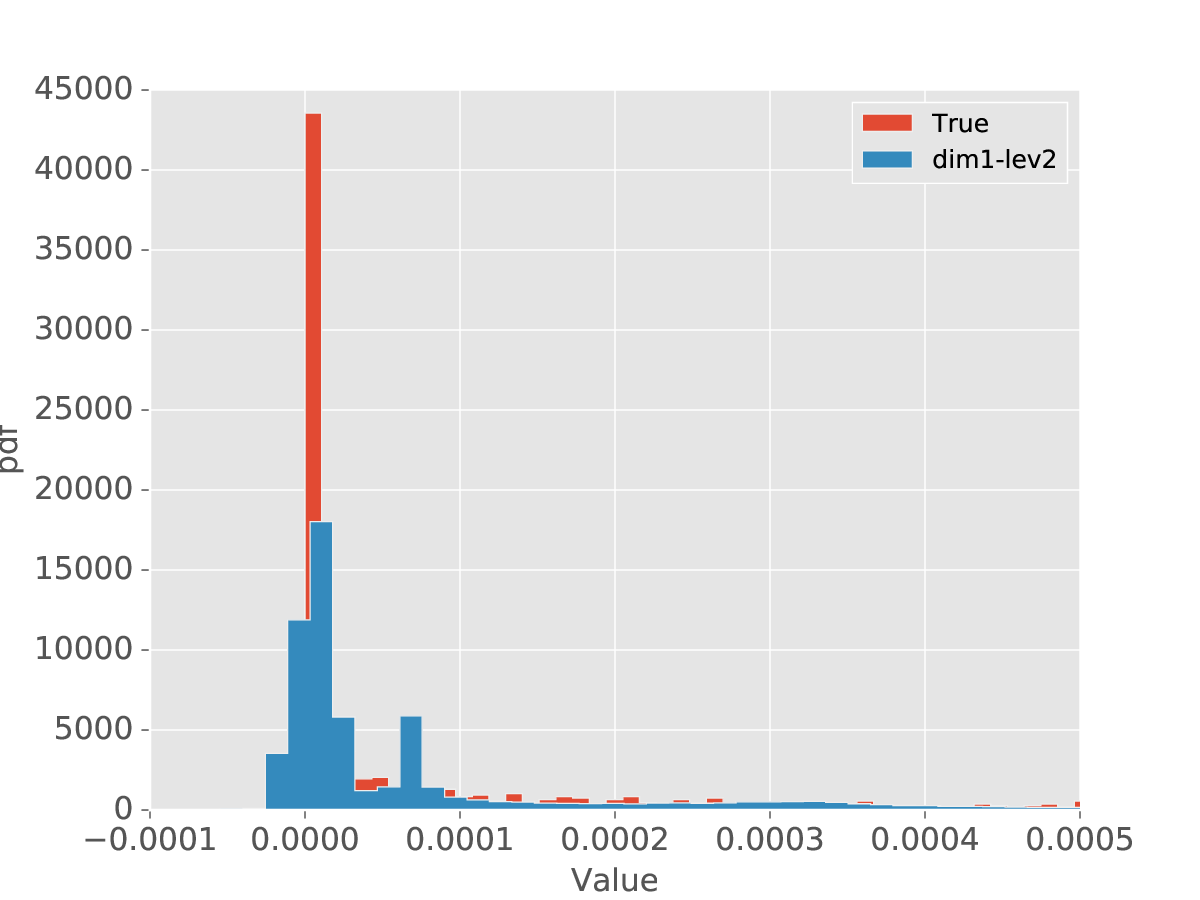, width=0.40\textwidth}
	\psfig{figure=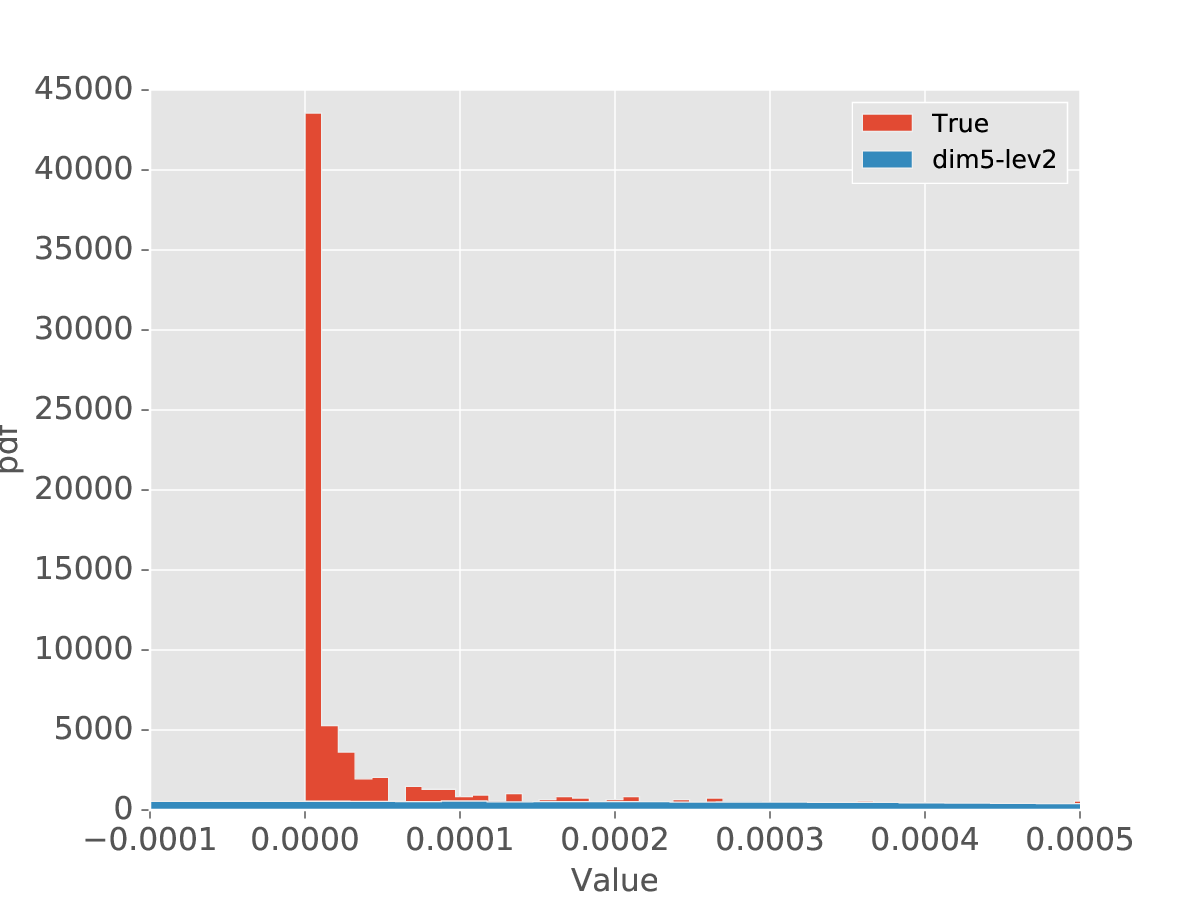, width=0.40\textwidth}
	}
\centerline{
	\psfig{figure=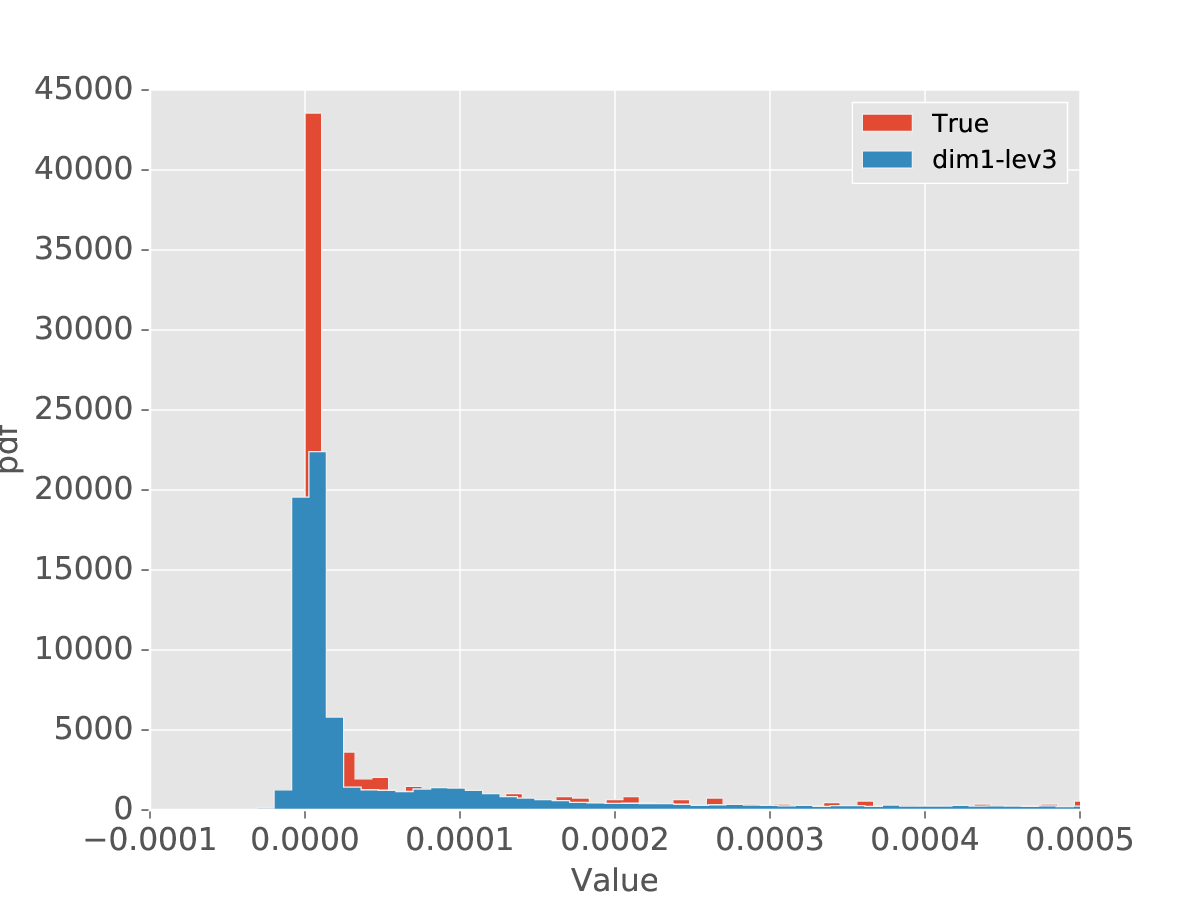, width=0.40\textwidth}
	\psfig{figure=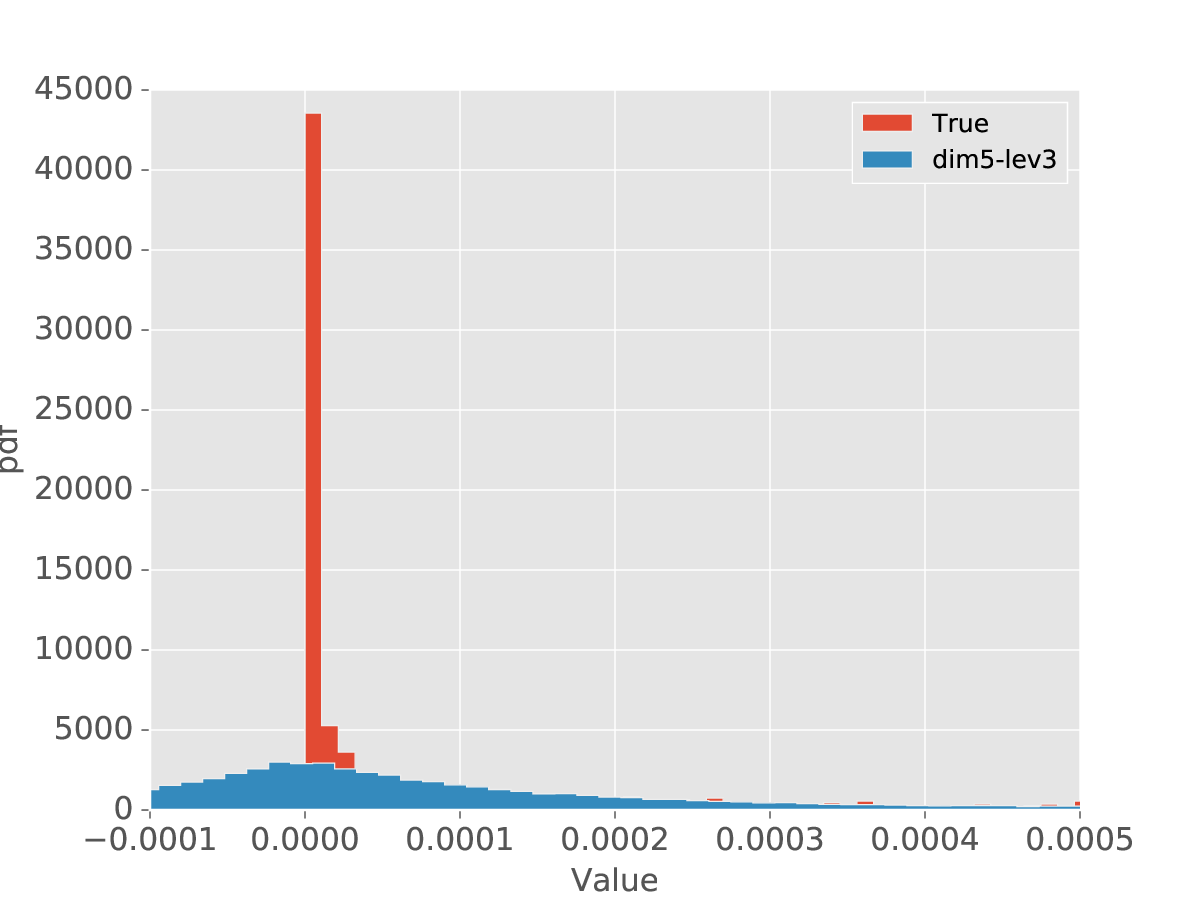, width=0.40\textwidth}
	}
\centerline{
	\psfig{figure=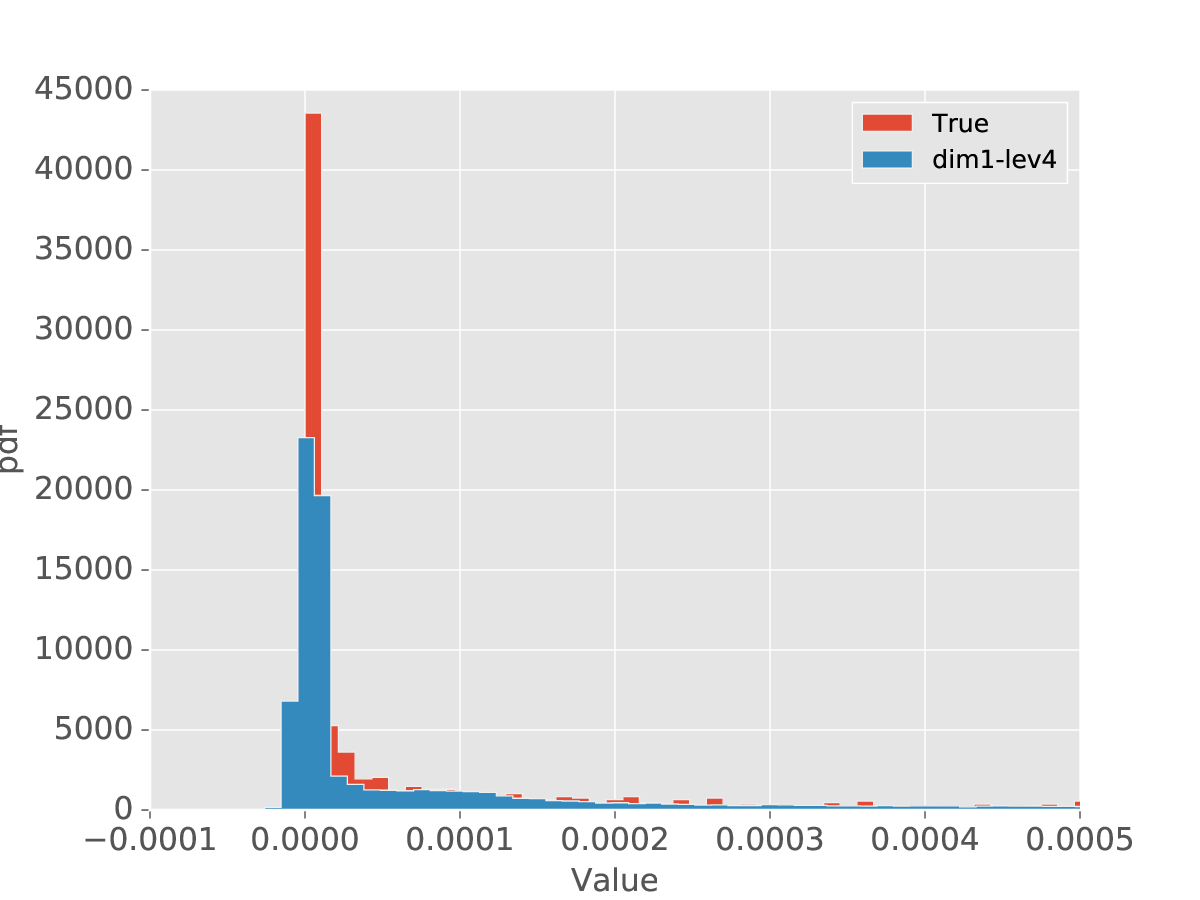, width=0.40\textwidth}
	\psfig{figure=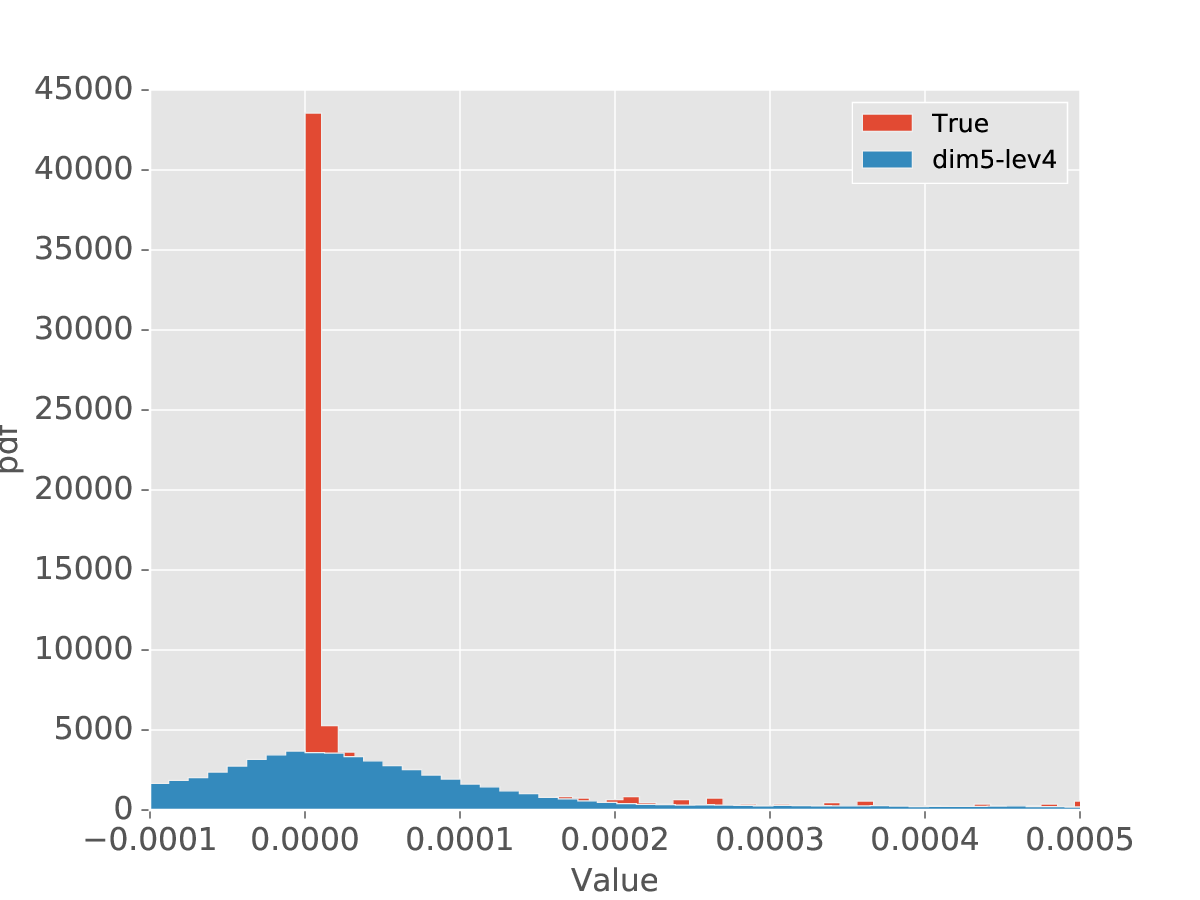, width=0.40\textwidth}
	}
\centerline{
	\psfig{figure=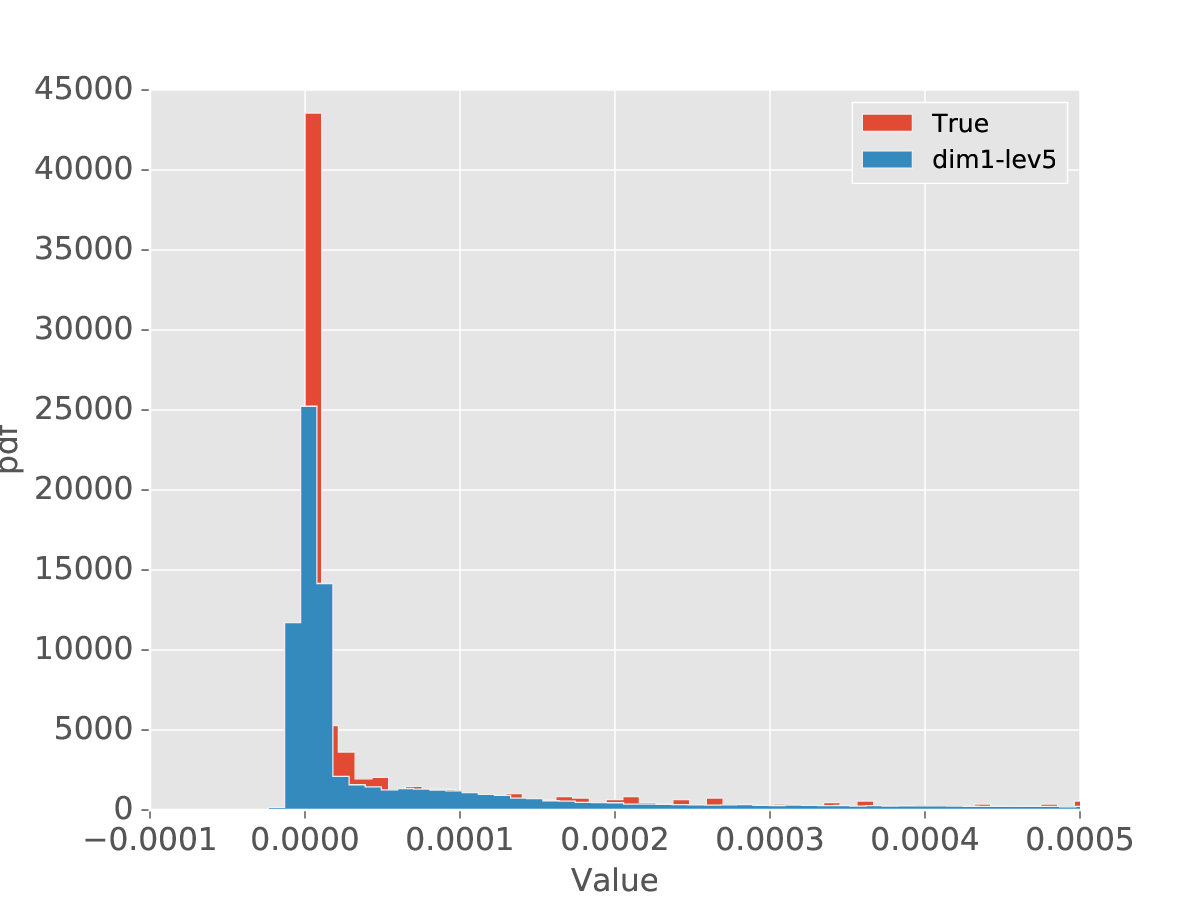, width=0.40\textwidth}
	\psfig{figure=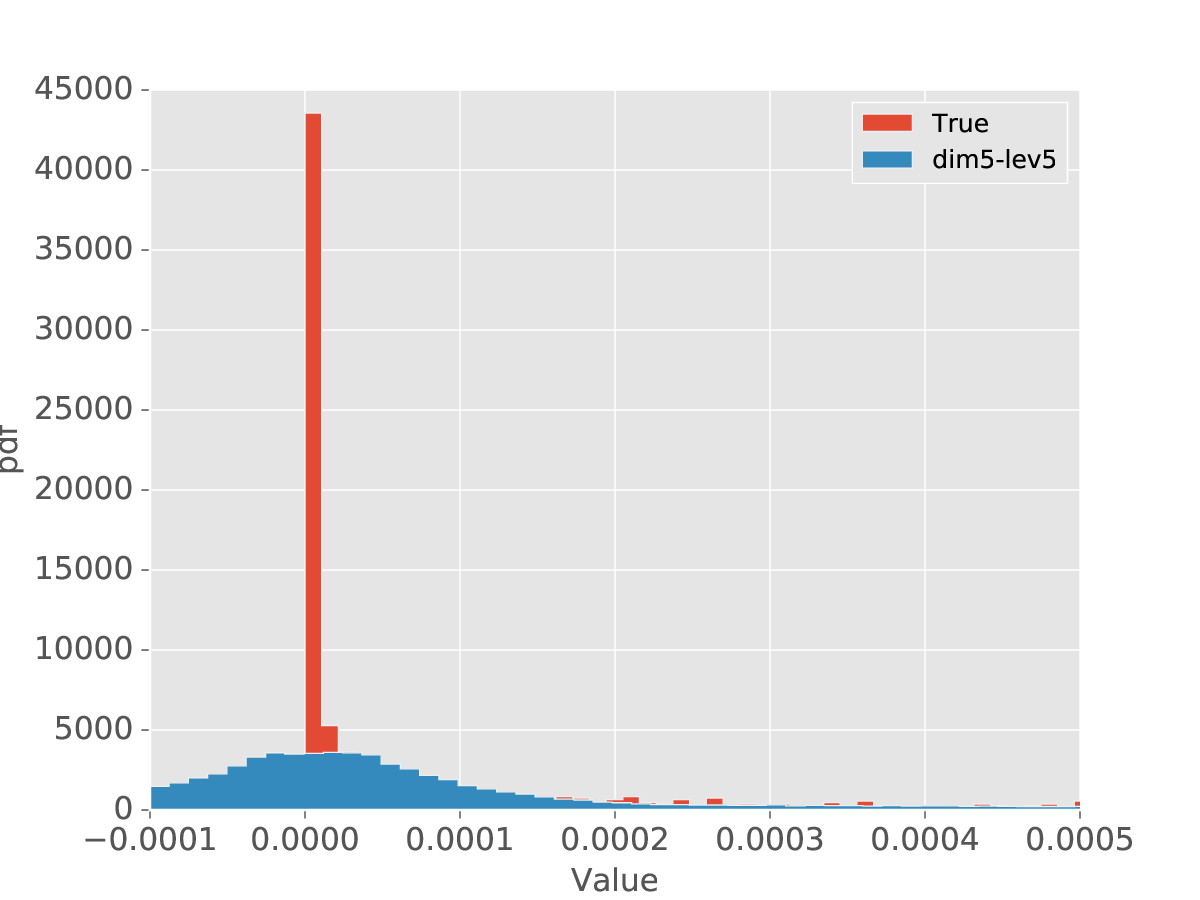, width=0.40\textwidth}
}
\caption{Histogram comparison based on MC samples from the 1d- (left column) and 5d- (right column) chaos expansions with that based on $1000$ MC samples drawn from the TOUGH2 simulator (all histograms are normalized such that the bins integrate to 1). \label{fig:histograms_comparison}}
\end{figure}

Finally, we compute the empirical pdfs based on $10^{5}$ MC samples of all
5-d and 1-d PC expansions that we have available and compare their
histograms with that of the $1000$ MC samples that are available
directly from the TOUGH2 model. The results are shown in
Fig. \ref{fig:histograms_comparison}. It is observed again that those
based on the 1-d expansions provide far better approximations of the true
histogram which is by definition bounded from below at $0$ while those
based on the full expansions fail dramatically to capture the lower tail
behavior. The computational resources required for the estimation of
the expansions described are also in favor of the basis reduction
methodology: The 1-d expansion based on an orthogonal projection $\bw$
that was computed using $\ell = 3$ quadrature rule required a total of
$241$ model evaluations for the estimation of the $3$rd order 5-d PC
plus $33$ model evaluations for a level-5 1-d quadrature rule
resulting in a total of $274$ model evaluations and eventually
provides a more accurate density function than a $\ell = 4$, or $5$ rule for the full PC
that require $801$ and $2433$ model evaluations respectively! 

%\begin{figure}
%\centering
%\includegraphics[width = 0.49\textwidth]{./images/QoI_pdf_dim1lev2.eps}
%\includegraphics[width = 0.49\textwidth]{./images/QoI_pdf_dim5lev2.eps}
%\includegraphics[width = 0.49\textwidth]{./images/QoI_pdf_dim1lev3.eps}
%\includegraphics[width = 0.49\textwidth]{./images/QoI_pdf_dim5lev3.eps}
%\includegraphics[width = 0.49\textwidth]{./images/QoI_pdf_dim1lev4.eps}
%\includegraphics[width = 0.49\textwidth]{./images/QoI_pdf_dim5lev4.eps}
%\includegraphics[width = 0.49\textwidth]{./images/QoI_pdf_dim1lev5.eps}
%\includegraphics[width = 0.49\textwidth]{./images/QoI_pdf_dim5lev5.eps}
%caption{Histogram comparison based on MC samples from the 1d- (left column) and 5d- (right column) chaos expansions with that based on $1000$ MC samples drawn from the TOUGH2 simulator. \label{fig:histograms_comparison}}
%\end{figure}

\section{Conclusions}

I have presented a methodology that combines the active subspace approach for dimensionality reduction with the construction of polynomial chaos surrogates for the purpose of efficient exploration of response surfaces and uncertainty propagation problems. This approach, serving at the same time as an extension of the basis adaptation framework \cite{tipireddy} for Homogeneous Chaos expansions, due to its applicability to generalized Polynomial Chaos, provides an explicit expression of the gradient covariance matrix in terms of the chaos coefficients that allows fast computation of the active subspace that is not only useful for dimension reduction, but even for sensitivity analysis purposes. Besides the attractive features in terms of computational efficiency and improved accuracy of the surrogate that were mentioned throughout the manucript and demonstrated in the numerical examples, the method shows the direction to future challenges: Precisely, the extension of the current approach to higher-dimensional active subspaces, either by finding {\color{blue}a} way to map the projected variables to a vector of independent ones or by constructing orthogonal polynomials with respect to arbitrary joint distributions, faces several obstacles as detailed in \cite{soize_desceliers} but it would provide a new ground on polynomial chaos-based dimensionality reduction methods.

\appendix

\section{Computation of the stiffness matrix $K_{ij}$}
\label{sec:stiff_comp}

First write the partial derivatives 
\begin{equation}
\frac{\partial \psi_{\balpha}(\bxi)}{\partial \xi_i} =
\frac{\partial}{\partial \xi_i} \prod_{m = 1}^d
\psi_{\alpha_m}(\xi_m) = \psi'_{\alpha_i}(\xi_i)
\prod_{\substack{m=1\\ m\neq i}}^d \psi_{\alpha_m}(\xi_m).
\end{equation}
Then for any $\balpha, \bbeta \in \calJ_P$, we have 
\begin{eqnarray}
\begin{array}{ccl}
\left(K_{ij}\right)_{\balpha \bbeta} & = & \displaystyle{ \E\left\{ \frac{\partial \psi_{\balpha}(\bxi)}{ \partial \xi_i} \cdot
  \frac{\partial \psi_{\bbeta}(\bxi)}{\partial \xi_j}\right\} } \\ & = &
\displaystyle{ \E\left\{ \left( \psi'_{\alpha_i}(\xi_i)
\prod_{\substack{m=1\\ m\neq i}}^d \psi_{\alpha_m}(\xi_m) \right) \cdot \left( \psi'_{\beta_j}(\xi_j)
\prod_{\substack{m=1\\ m\neq j}}^d \psi_{\beta_m}(\xi_m)\right)
\right\} }
\end{array}
\end{eqnarray}
which for $i \neq j$ gives
\begin{eqnarray}
\begin{array}{ccl}
\left( K_{ij}\right)_{\balpha \bbeta} & = & \displaystyle{\left(
\prod_{\substack{m = 1\\ m \neq i, m\neq j}}^d \E\left\{
  \psi_{\alpha_m}(\xi_m)\psi_{\beta_m}(\xi_m)\right\} \right) \cdot \E\left\{
  \psi'_{\alpha_i}(\xi_i) \psi_{\beta_i}(\xi_i)\right\} \cdot
\E\left\{\psi_{\alpha_j}(\xi_j) \psi'_{\beta_j}(\xi_j)\right\} } \\ & =
& \displaystyle{ \left( \prod_{\substack{m = 1\\ m \neq i, m\neq j}}^d 
  \delta_{\alpha_m, \beta_m}  \right) \cdot \E\left\{
  \psi'_{\alpha_i}(\xi_i) \psi_{\beta_i}(\xi_i)\right\} \cdot
\E\left\{\psi_{\alpha_j}(\xi_j) \psi'_{\beta_j}(\xi_j)\right\} 
}
\end{array}
\end{eqnarray} 
and for $i = j$ gives 
\begin{eqnarray}
\begin{array}{ccl}
\left( K_{ii}\right)_{\balpha \bbeta} & = & \displaystyle{\left(
\prod_{\substack{m = 1\\ m \neq i}}^d \E\left\{
  \psi_{\alpha_m}(\xi_m)\psi_{\beta_m}(\xi_m)\right\} \right) \cdot \E\left\{
  \psi'_{\alpha_i}(\xi_i) \psi'_{\beta_i}(\xi_i)\right\} } \\ & =
& \displaystyle{ \left( \prod_{\substack{m = 1\\ m \neq i}}^d 
  \delta_{\alpha_m, \beta_m}  \right) \cdot \E\left\{
  \psi'_{\alpha_i}(\xi_i) \psi'_{\beta_i}(\xi_i)\right\} .
}
\end{array}
\end{eqnarray}
The computation of the terms $ \E\left\{
  \psi'_{\alpha}(\xi) \psi_{\beta}(\xi)\right\}$, and $\E\left\{
  \psi'_{\alpha}(\xi) \psi'_{\beta}(\xi)\right\} $ depends on
the choice of polynomials. The case of Hermite polynomials was derived in \cite{yang_lin}. Here I explore the case of Legendre polynomials.

%\subsection{Hermite Chaos}

%If $\psi_\alpha(\xi) = \frac{h_\alpha(\xi)}{||h_\alpha(\xi)||}$ are the
%normalized Hermite polynomials, we have that 
%\begin{equation}
%\psi'_{\alpha}(\xi) = \sqrt{\alpha} \psi_{\alpha-1}(\xi), \ \ \alpha
%\geq 1
%\end{equation}
%and $\psi'_0(\xi) = 0$ which gives
%\begin{equation}
%\E\left\{
%  \psi'_{\alpha}(\xi) \psi_{\beta}(\xi)\right\} =
%\sqrt{\alpha}\E\left\{\psi_{\alpha-1}(\xi)\psi_{\beta}(\xi)\right\} =
%\sqrt{\alpha} \delta_{\alpha-1, \beta}
%\end{equation}
%and 
%\begin{equation}
%\E\left\{ \psi'_{\alpha}(\xi) \psi'_{\beta}(\xi)\right\} = \sqrt{\alpha\beta}
%\E\left\{\psi_{\alpha-1}(\xi)\psi_{\beta-1}(\xi)\right\} = \sqrt{\alpha\beta}\delta_{\alpha-1,\beta-1}.
%\end{equation}

%\subsection{Legendre Chaos}
In the case where $\psi_\alpha(\xi) =
\frac{\ell_\alpha(\xi)}{||\ell_{\alpha}(\xi)||}$ are the normalized Legendre polynomials, the expectations in the last
two terms of the above relation can be
computed recursively using the relation
\begin{equation}
(2\alpha + 1) \ell_{\alpha}(\xi) = \frac{d}{d \xi}
\left[ \ell_{\alpha + 1}(\xi) - \ell_{\alpha - 1}(\xi)\right]
\end{equation}
that gives 
\begin{eqnarray}
\label{eq:recur_1}
\begin{array}{ccl}
\E\left\{ \ell'_{\alpha}(\xi) \ell_{\beta}(\xi)\right\} & = & \displaystyle{
\delta_{\alpha-1, \beta} +
\E\left\{\ell'_{\alpha-2}(\xi)\ell_{\beta}(\xi)\right\} }\\ & = & \displaystyle{
\delta_{\alpha-1, \beta} + \delta_{\alpha - 3, \beta} +
\delta_{\alpha-5, \beta} + \dots} .
\end{array}
\end{eqnarray}
and 
\begin{eqnarray}
\label{eq:recur_2}
\begin{array}{ccl}
\displaystyle {\E\left\{ \ell'_{\alpha}(\xi)
    \ell'_{\beta}(\xi)\right\}} & = & \displaystyle{
(2\alpha-1)\delta_{\alpha-1, \beta-1} +
(2\alpha-1)\E\left\{\ell_{\alpha-1}(\xi) \ell'_{\beta-2}(\xi)\right\} }
 \\ & & \displaystyle{ + (2\beta-1) \E\left\{
  \ell'_{\alpha-2}(\xi)\ell_{\beta-1}(\xi)\right\} +
\E\left\{\ell'_{\alpha-2}(\xi)\ell'_{\beta-2}(\xi)\right\} }\\ & = &  \displaystyle{
(2\alpha - 1) \left(\delta_{\alpha-1,\beta-1}  + \delta_{\alpha-1,
    \beta-3} + \dots\right) } \\ & & \displaystyle{ + (2\alpha - 5)
\left(\delta_{\alpha -3, \beta-1} + \delta_{\alpha -3, \beta - 3} +
  \dots \right) } \\
  & & \displaystyle{ + (2\alpha - 9)\left(\delta_{\alpha-5, \beta- 1} +
    \delta_{\alpha - 5, \beta- 3} + \dots \right) + \dots .}
\end{array}
\end{eqnarray}

Using eqs. (\ref{eq:recur_1}) and (\ref{eq:recur_2}) finally I compute 
\begin{eqnarray}
\E\left\{ \psi'_{\alpha}(\xi) \psi_{\beta}(\xi)\right\}  =
\sqrt{(2\alpha+1)(2\beta+1)} \E\left\{ \ell'_{\alpha}(\xi) \ell_{\beta}(\xi)\right\}
\end{eqnarray}
and
\begin{equation}
\E\left\{\psi'_\alpha(\xi) \psi'_\beta(\xi)\right\} =
\sqrt{(2\alpha+1)(2\beta+1)} \E\left\{\ell'_{\alpha}(\xi) \ell'_{\beta}(\xi)\right\}.
\end{equation}

%for $\alpha, \beta > 1$ we define $\gamma = \min(\alpha, \beta)$
%and 
%\begin{eqnarray}
%C = \left\{\begin{array}{ll} 0, & \gamma = 2k \\
 %   \E\left\{\psi'_1(\xi)\psi_{\max(\alpha, \beta)}'(\xi)\right\}, & \gamma = 2k+1\end{array}\right.
%\end{eqnarray}
%and we have
%\begin{eqnarray}
%\begin{array}{ccl}
%\displaystyle { \E\left\{\psi'_\alpha(\xi) \psi'_\beta(\xi)\right\}}
%& = & \displaystyle { \sum_{i=0}^{k-1}
%(2\alpha-4i-1)(2\beta - 4i -1)\delta_{\alpha-2i-1, \beta-2i -1} } \\ &
%& \displaystyle{ +
%\sum_{i=0}^{k-1} (2\alpha - 4i -1)\E\left\{\psi_{\alpha -2i -1}(\xi)\psi'_{\beta -2i
 %   -2}(\xi)\right\} } \\ & & \displaystyle{ + \sum_{i=0}^{k-1} (2\beta - 4i -1)\E\left\{\psi'_{\alpha -2i -2}(\xi)\psi_{\beta -2i
%    -1}(\xi)\right\}  } + C.
%\end{array}
%\end{eqnarray}

%%%%%%%%%%%%%%%%%%%%%%%%%%%%%%%%%%%%%%%%%%%%%%%%%%%%%%%%%%%%%%%%%%%%%%
% The bibliography is stored in an external database file
% in the BibTeX format (file_name.bib).  The bibliography is
% created by the following command and it will appear in this
% position in the document. You may, of course, create your
% own bibliography by using thebibliography environment as in
%
% \begin{thebibliography}{12}
% ...
% \bibitem{itemreference} D. E. Knudsen.
% {\em 1966 World Bnus Almanac.}
% {Permafrost Press, Novosibirsk.}
% ...
% \end{thebibliography}

% Here's where you specify the bibliography style file.
% The full file name for the bibliography style file 
% used for an ASME paper is asmems4.bst.
\bibliographystyle{asmems4}

% Here's where you specify the bibliography database file.
% The full file name of the bibliography database for this
% article is asme2e.bib. The name for your database is up
% to you.
\bibliography{references}

\end{document}